\documentclass[acmsmall]{acmart}

\AtBeginDocument{%
  \providecommand\BibTeX{{%
    \normalfont B\kern-0.5em{\scshape i\kern-0.25em b}\kern-0.8em\TeX}}}

\setcopyright{acmcopyright}
\copyrightyear{2018}
\acmYear{2018}
\acmDOI{XXXXXXX.XXXXXXX}

\acmJournal{JACM}
\acmVolume{37}
\acmNumber{4}
\acmArticle{111}
\acmMonth{8}

\usepackage[framemethod=tikz]{mdframed}
\usepackage{xcolor, soul}  
\usepackage{xcolor}

\mdfdefinestyle{mpdframe}{
    frametitlebackgroundcolor   =black!15,
    frametitlerule              =true,
    roundcorner                 =5pt,
    middlelinewidth             =1pt,
    innermargin                 =0.3cm,
    outermargin                 =0.3cm,
    innerleftmargin             =0.3cm,
    innerrightmargin            =0.3cm,
    innertopmargin              =0.5cm,
    innerbottommargin           =0.5cm
}

\begin{document}

\def \rqa {How sensitive is TFix to potential domain shifts?}
\def \rqb {How effective, practical, and reliable is the proposed DA framework for TFix?}
\def \rqbw {How effectively does the proposed DA framework improve the accuracy of TFix on the target projects?}
\def \rqbx {How practical is the proposed DA framework for TFix in terms of efficiency?}
\def \rqby {How reliable is the proposed DA framework for TFix in terms of vulnerability to exposure bias?}
\def \rqc {How effectively does the proposed data synthesis method improve TFix on projects with no labeled data?}
\def \rqcw {How effectively does TBug synthesize buggy code lines?}
\def \rqcx {What is the improvement of using the synthetic dataset of TBug in the DA framework when used for the target projects with no labeled data?}
\def \rqd {Do the methods and findings of this study extend to CodeXGLUE?}
\def \rqdw {How vulnerable is CodeXGLUE to the threat of potential domain shifts?}
\def \rqdx {What is the benefit of the proposed DA framework with \full{} on CodeXGLUE?}
\def \rqdy {How effectively does the unsupervised DA framework improve CodeXGLUE using CodeXBUG?}

\def \full {\textit{FullFineTuning}}
\def \fullm {FFT}
\def \adapter {\textit{TuningWithLightWeightAdapterLayers}}
\def \adapterm {TLWAL}
\def \curr {\textit{CurriculumLearning}}
\def \currm {CL}
\def \currl {\textit{CurriculumLearningWithLength}}
\def \currlm {CLL}
\def \currc {\textit{CurriculumLearningWithConfidence}}
\def \currcm {CLC}
\def \currs {\textit{CurriculumLearningWithSimilarity}}
\def \currsm {CLS}

\title{Improving Automated Program Repair with Domain Adaptation}

\author{Armin Zirak}
\email{armin.zirak@ucalgary.ca}
\orcid{1234-5678-9012}
\author{Hadi Hemmati}
\email{hadi.hemmati@ucalgary.ca}
\affiliation{%
  \institution{University of Calgary}
  \city{Calgary}
  \state{Alberta}
  \country{Canada}s
}
s

\begin{abstract}
Automated Program Repair (APR) is defined as the process of fixing a bug/defect in the source code, by an automated tool. APR tools have recently experienced promising results by leveraging state-of-the-art Neural Language Processing (NLP) techniques. APR tools such as TFix and CodeXGLUE that combine text-to-text transformers with software-specific techniques are outperforming alternatives, these days. However, in most APR studies the train and test sets are chosen from the same set of projects (i.e., when APR fixes a bug in the test set from project A, the model has already seen example fixed bugs from project A in the training set). In the real world, however, APR models are meant to be generalizable to new and different projects. Therefore, there is a potential threat that reported APR models with high effectiveness perform poorly when the characteristics of the new project or its bugs are different than the training set's (``Domain Shift'').

In this study, we first define the problem of domain shift in automated program repair. Next, we measure the potential damage of domain shift on two state-of-the-art APR models (TFix and CodeXGLUE). Based on this observation, we then propose a domain adaptation framework that can adapt an APR model for a given target project. We conduct an empirical study with three domain adaptation methods \full{}, \adapter{}, and \curr{} and two APR models on 611 bugs from 19 projects. 

The results show that our proposed framework on average can improve the effectiveness of TFix by 13.05\% and CodeXGLUE by 23.4\%, in terms of ``Exact Match''. Through experiments, we also show that the framework provides high efficiency and reliability (in terms of ``Exposure Bias''). 

Another contribution of this study is the proposal of a data synthesis method to address the lack of labeled data in APR (bug-fix pairs). We leverage transformers to create a bug generator model. We use the generated synthetic data to domain adapt TFix and CodeXGLUE on the projects with no data (Zero-shot learning), which results in an average improvement of 5.76\% and 24.42\% for TFix and CodeXGLUE, respectively.
  
\end{abstract}



\maketitle

\section{Introduction}
One of the most ideal automated software engineering tasks that can potentially reduce software development/maintenance costs, significantly, is called Automated Program Repair (APR) \cite{gazzola2017automatic}.
APR is defined as the process of fixing a bug/defect in the source code by an automated tool. Traditional APR techniques use rule-based or search-based methods. For example, they apply a set of predefined modifications on the buggy code until it passes the test cases. 
Modern APR techniques, however, are set to be more generic \cite{tufano2019empirical}. These models are primarily learning-based and inspired by Natural Language Processing (NLP) research \cite{berabi2021tfix, chen2019sequencer, lutellier2020coconut, tufano2019empirical, tufano2018empirical, bader2019getafix, lu2021codexglue}.

Recent APR studies leverage Neural Machine Translation (NMT) techniques. They define APR as a \textit{translation task} that translates a buggy code line to a fixed code line. This led to the emergence of NMT-based APR models, such as SequenceR \cite{chen2019sequencer}, CoCoNuT \cite{lutellier2020coconut}, CodeXGLue \cite{lu2021codexglue}, and TFix \cite{berabi2021tfix}. The underlying architecture of these techniques are based on sequence-to-sequence models such as LSTM\cite{li2022dear, chen2019sequencer}, Encoder-Decoder \cite{lu2021codexglue, tufano2019empirical}, and Transformers \cite{berabi2021tfix}.

Learning-based APR models usually perform better than traditional search-based and rule-based techniques. They show more promising results and fix a diverse range of errors. However, they potentially suffer from a major issue; Domain Shift (DSH) \cite{redko2019advances}. Domain shift arises when a model is trained on a dataset (source dataset) and then is tested on a different dataset (target dataset), where the distribution of the target dataset differs from the source dataset. This might cause a considerable drop in the model's performance \cite{redko2019advances, amodei2016concrete, sun2016return}. 

The problem of domain shift has been studied in Machine Learning and Natural Language Processing, for many years. Several studies proposed domain adaptation methods. These methods either make models resistant to domain shift or adapt them to the target domain. \cite{wang2018deep, patel2015visual, farahani2021brief, saunders2021domain}.
Within the domain of software engineering, a couple of studies have analyzed the effect of transferring models from one domain to another. They measured the drops and tried to find solutions. For example, some transfer models trained on synthetic data to real-world data \cite{he2022distribution}. Some transfer models trained on languages with highly available resources such as Java to adapt to low-resource languages such as R \cite{salza2021effectiveness, chai2022cross}. Finally, cross-project domain shift is the primary attention of some studies. They transfer models trained on one or a couple of projects to different projects \cite{jin2021cross,ma2012transfer, limsettho2018cross, liu2019two, de2021comparing, liu2020cd, jin2021cross}. 

In this paper, we argue that, similar to other tasks in software engineering, APR models are in the risk of domain shift. APR studies are typically designed in a way that the models are trained and tested on the same set of projects. That means the model might perform poorly on new projects which are not part of the training set. This is unfortunate because in reality the likely scenario of a practical usage of an APR model will be applying it on a real-world project (e.g., a company's project) which is not part of the open source benchmark training data for APR studies. Our study shows that, for example, domain shift reduces the accuracy of TFix (Large architecture) and CodeXGLUE (Small dataset) 11.82\% and a 26.13\%, respectively, when tested on on-seen projects. These examples motivate us to leverage domain adaptation methods to maintain the accuracy of APR models on target projects.


In our experiments, we choose TFix \cite{berabi2021tfix} and CodeXGLUE \cite{lu2021codexglue} as the state-of-the-art APR models for this study. TFix \cite{berabi2021tfix} is a successful APR model that outperforms many of the previously proposed methods by a convincing margin\cite{berabi2021tfix}. It leverages a novel NMT method, Text-to-Text T5 transformers \cite{raffel2020exploring}, and augment it with a software-specific input (based on static analysis). TFix fixes generic JavaScript buggy programs. CodeXGLUE is another effective APR method that uses a well-known large pre-trained language model, CodeBERT \cite{feng2020codebert}. CodeXGLUE is tested on two datasets of Java function definitions.

We use two architecture sizes of TFix; Large and Small. The large size is the primary one with the highest effectiveness. On the other hand, small is more efficient. Also, the small size has fewer learnable parameters. This might cause a different reaction to domain shift and domain adaptation. We also use two different datasets of CodeXGLUE; Small and Medium. The small dataset has shorter and easier to fix code lines. We use the three architectures on 611 bugs, 19 projects, and three datasets. 

We analyze domain shift and domain adaptation in the context of cross-project APR. We start with defining \textit{Domain Shift in Automated Program Repair}. Next, we measure the damage of domain shift to TFix and CodeXGLUE. We design two experiments to compare the case in which domain shift occurs with the case where the test domain does not shift. We show a substantial drop of 11.82\% in TFix-Large and a small drop of 1.47\% in TFix-Small. We also show that CodeXGLUE's accuracy drops 26.13\% in the small dataset and 5.32\% in the medium dataset. These all confirm that APR methods are sensitive to domain shift and further motivates us to employ domain adaptation methods to address it.

Accordingly, we propose a domain adaptation framework that adapts an APR model to a target project. We explore three well-known NLP DA methods (\full{}, \adapter{} and \curr{}) that can fit into our framework. We do an empirical study of all the mentioned methods on TFix. This results in an improvement of 13.05\% and 6.66\% for TFix-Large and TFix-Small using \full{}. We show that the proposed methods are efficient. Adapted models can be created in a feasible time with low resource requirements. We show that the results are reliable by measuring the exposure bias effect. Furthermore, we validate the findings for CodeXGLUE on both small and medium datasets. We use \full{} as the most effective, efficient and reliable DA method to adapt CodeXGLUE on target projects. We report an improvement of 23.42\%  in the small dataset, and 4.26\% in the medium datasets. 

Given that our framework is supervised, it requires a clean labeled dataset for each target project. This is a problematic assumption as many software engineering projects may come with no history; even otherwise, creating a clean and large-enough dataset (buggy code snippets and their fixed versions) is not trivial. Mapping a buggy code line to its fixed version (which must be independent of the other code changes) requires bug/fix matching algorithms. The existing bug/fix matching algorithms filter out a significant part of the change histories and keep only a few of them. This results in small curated datasets for each target project of interest \cite{tufano2019empirical, berabi2021tfix}.

Therefore, we propose a data synthesis method to create labeled APR datasets. Inspired by NMT-based APR models, we define \textit{bug synthesis} as the task of translating non-buggy source code to buggy source code. We introduce a bug generator model that injects a bug to a code line. Our bug generator model is based on NMT architectures. For example, we take the T5 and CodeBERT-Decoder models of TFix and CodeXGLUE and finetune them on the ``reversed datasets''. We call them TBug and CodeXBUG. That means TBug and CodeXBUG take a non-buggy code line and inject a bug into it. Our results show 41.4\% and 44.68\% effectiveness for small and large sizes of TBug. We show that using the synthetic bug datasets of TBug-Small and TBug-Large leads to the improvement of 4.68\% and 5.76\% in DA. We also show that using synthetic bugs of CodeXBUG makes an improvement of 23.42\% and 4.32\% on the small and medium datasets. Note that the usage of these synthetic datasets are not limited only to the APR problem but can also be used in various other software engineering tasks such as fault localization and test generation.


The following is the complete list of research questions we answer in this study.

\begin{itemize}
\item \textbf{RQ1:} \rqa{}
\item \textbf{RQ2:} \rqb{}
\begin{itemize}
\item \textbf{RQ2.1:}\rqbw{}
\item \textbf{RQ2.2":}\rqbx{}
\item \textbf{RQ2.3:}\rqby{}
\end{itemize}
\item \textbf{RQ3:} \rqc{}
\begin{itemize}
\item \textbf{RQ3.1:} \rqcw{}
\item \textbf{RQ3.2:} \rqcx{}
\end{itemize}
\item \textbf{RQ4:} \rqd{}
\begin{itemize}
\item \textbf{RQ4.1:} \rqdw{}
\item \textbf{RQ4.2:} \rqdx{}
\item \textbf{RQ4.3:} \rqdy{}
\end{itemize}
\end{itemize}

The rest of this study is organized as follows. First, we explain the background of the work in \ref{sc:background}. We propose our domain adaptation framework and data synthesis method in \ref{sc:method}. We explain the experiment design in \ref{sc:experiments} and report the results in \ref{sc:evaluation}. We walk through the related work in \ref{sc:related_work} and explain the advantages of our study. We specify the potential directions for future studies in \ref{sc:future}. Finally, we conclude the study in \ref{sc:conclusion}.

\vspace{5mm}
\noindent{}We publish all the source codes, models and results in a public repository.\footnote{https://github.com/arminzirak/TFix}

\section{Background}
\label{sc:background}
In this section, we describe the background and history of Automated Program Repair (APR). We walk through many APR techniques. Then, we explain TFix and CodeXGLUE (the selected models for this study) in more detail. Finally, we describe the general Domain Adaptation (DA) in Neural Machine Translation (NMT), its important methods, and evaluation criteria.

\subsection{Automated Program Repair}
Automated Program Repair (APR) is the task of accepting a buggy code snippet as the input and returning the corresponding patch (fixed line(s)) as the output. Most APRs have limited themselves to one-line patches, where the bug can be fixed by only editing one line of code. A \textit{buggy code line} is defined as a code line that raises compilation errors, fails test cases, or triggers error detectors. The failures not captured by compilers or parsers are called ``logical errors''. Logical errors are a more challenging area of APR and attract the focus of novel APR methods \cite{chen2019sequencer, tufano2018empirical, tufano2019empirical, lutellier2020coconut, lu2021codexglue, berabi2021tfix, bader2019getafix}. Therefore, we also narrow the scope of this paper to logical one-line errors. A fixed code line (patch) is referred to as the line that a developer replaces and no longer has that failure. APR methods assume that buggy lines are detected and provided beforehand. Detection (testing) and localization of buggy code lines (fault localization) are separate lines of research in software engineering that are outside of the scope of this paper.

\label{APR_bg}
APR has been a classic line of research in software engineering \cite{gazzola2017automatic, monperrus2018automatic}. A large group of traditional APR methods explored search-based software engineering \cite{le2012systematic, martinez2016astor, qi2014strength, le2011genprog}. 
%
Learning-based approaches grew as a new generation of APR techniques. They \textit{learn} code transformation from the history of human-made changes such as version control commits. Some studies formulate APR as a sequence modeling task and use neural networks to address it. \cite{bhatia2018neuro, gupta2017deepfix, allamanis2017learning, vasic2019neural, hellendoorn2019global}. However, these methods are mostly limited to specific bug types or datasets and do not generalize to various general logical errors. \cite{sobreira2018dissection}

Most recently APR has been defined as a \textit{translation task} of translating buggy code lines to non-buggy code lines \cite{tufano2018empirical,tufano2019empirical}. These techniques leveraged an NMT method, Encoder-Decoder, on processed AST trees to fix bugs. \citeauthor{chen2019sequencer}
SequenceR \cite{chen2019sequencer}, DLFix \cite{li2020dlfix}, CoCoNut \cite{lutellier2020coconut}, Hoppity \cite{dinella2020hoppity}, and GetaFix \cite{bader2019getafix} are all examples of NMT-based  APRs in the literature. TFix \cite{berabi2021tfix} is a novel NMT-based APR that uses Transformers architecture and outperforms alternatives by a considerable margin. We explain TFix, in details, in the next section.


In a separate line of research, large language models from NLP have been adopted for source code representation (code embedding). Consequently, several software engineering tasks, including APR, have leveraged such models. \citeauthor{feng2020codebert} introduced CodeBERT \cite{feng2020codebert}, a bimodal pretrained model for programming languages that learns general purpose representation of source code. Several studies used CodeBERT to propose APR methods \cite{kanade2020learning, chen2021plur, guo2020graphcodebert, lu2021codexglue}. CodeXGLUE is an example study by \citeauthor{lu2021codexglue} \cite{lu2021codexglue}. It  adds a decoder to CodeBERT and fine tunes it on an APR dataset of Java function definitions. 

We explain TFix and CodeXGLUE in the following two sections and justify why they are used as APR methods of this study.

\begin{figure}
  \centering
  \includegraphics[width=\linewidth]{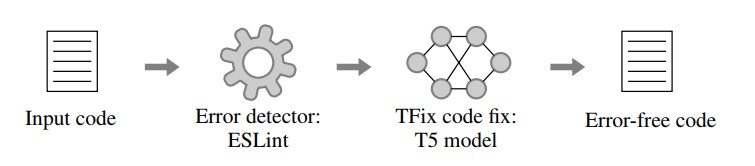}
  \caption{Pipeline of TFix.}
  \label{fig:tfix}
\end{figure}

\subsubsection{TFix}
The entire pipeline of TFix is shown in Figure \ref{fig:tfix}. First, TFix feeds buggy code to ESLint~\cite{eslint}. ESLint is an error detector tool that runs static algorithms on JS snippets. It extracts information about the buggy code line, such as error location, error type, and error message. A sample of Eslint output is as follows: \\

\fbox{%
\parbox{370pt}{%
  \textbf{line 670 error:} \fcolorbox{pink}{white}{guard-for-in} \\
  \textbf{error message:} \colorbox{pink}{the body of a for-in should be wrapped in an if statement to filter.}
    }%
}

\vspace{5mm}

The buggy code line, its surrounding context and the output of the ESLint are fed to the TFix model. The TFix model fixes the bug and returns the error-free code. The following shows the input and output formatting:

\vspace{1mm}
\begin{center}fix \fcolorbox{pink}{white}{error type} \colorbox{pink}{error message} \textcolor{red}{buggy line}: \fcolorbox{pink}{white}{error context}  $\Longrightarrow$ \textcolor{teal}{fixed line} \end{center} 
\vspace{1mm}

\citeauthor{berabi2021tfix} created a high-quality JavaScript(JS) program repair dataset of parallel data to train TFix. They collected 5.5 million GitHub commits and extracted $\sim100k$ aligned pairs of buggy code lines and their fixes. The extraction process invoked a static analyzer (ESLint) and a combination of greedy bipartite matching and Myers Diff algorithms. This resulted in a natural dataset from real bugs. Then, they finetuned \underline{T5} on the dataset according to the explained formatting. They called the resulting model \underline{TFix}. 

\citeauthor{berabi2021tfix} \cite{berabi2021tfix} reported TFix is more effective than alternatives including SequenceR, CoCoNuT, and Hoppity \cite{berabi2021tfix}. TFix is not limited to specific error types. It generates fixes with all types of errors detected by ESlint. It is tested on real data rather than synthetic data. In addition to the buggy line, it leverages ``code context''. Code context refers to the two neighboring lines of the buggy line. The code context provides the model with more information to fix the error. It also leverages the metadata of the static analyzer. The metadata consists of error type and error message. These all provide the model with more information about the bug. The method is based on the most state-of-the-art NLP and machine translation models; transformers \cite{barrault2020findings}. It also leverages the natural language information of T5. Although the method is relatively recent, it has been widely referenced recently. Code and dataset of TFix are released in public as open-source. Therefore, we choose TFix as one of the methods of this study. We analyze the threats of domain shift on it and address them using domain adaptation.

On the other hand, The main limitations of TFix is that first, it is examined only on a JS dataset and second, it requires the information of ESLint (and thus limited to bugs that ESLint can detect). Therefore, we also use CodeXGLUE as a different APR method to expand our discussions and add to the generalizability of findings.
 
\vspace{1mm}
\subsubsection{CodeXGLUE}
\citeauthor{lu2021codexglue} \cite{lu2021codexglue} introduced CodeXGLUE as a machine learning benchmark. They used BERT-style \cite{devlin2018bert}, GPT-style \cite{brown2020language}, and Encoder-Decoder models to solve 10 software engineering tasks on 14 datasets. APR is one of the tasks in their study. They propose a method based on CodeBERT that performs on a program repair dataset of java functions. Each instance includes a buggy java function definition and its fixed version. In the rest of this study, we refer to their APR method as ``CodeXGLUE''.

\begin{figure}[h]
  \centering
  \includegraphics[width=\linewidth]{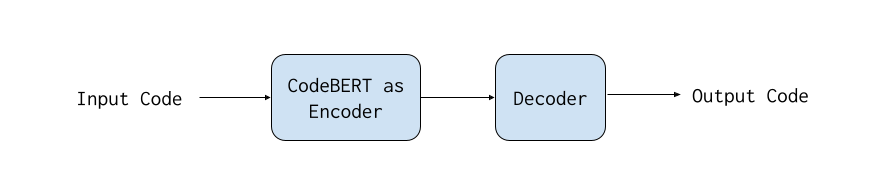}
  \caption{Encoder-Decoder Architecture of CodeXGLUE}
  \label{fig:codexglue}
\end{figure}

The high-level architecture of CodeXGLUE is depicted in Figure \ref{fig:codexglue}. It is an Encoder-Decoder framework. CodeBERT is used as the encoder, and a randomly initialized transformer with 6 layers, 768-dimensional hidden states, and 12 attention heads is used as the decoder. We call this architecture ``CodeBERT-decoder'' in the rest of this study.

CodeBERT is a bimodal pretrained model based on a transformer with 6 layers, 768-dimensional hidden states, and 12 attention heads. It is based on both programming language (PL) and natural language(NL)\cite{feng2020codebert}. \citeauthor{feng2020codebert} \cite{feng2020codebert} pretrained CodeBERT by ``masked language modeling'' and ``replaced token detection'' objectives. They used CodeSearchNet \cite{husain2019codesearchnet} as the training dataset which includes 2.4M functions with document pairs for six programming languages. 

\citeauthor{tufano2019empirical} collected a dataset of java functions from public Github events and selected bug-fix instances based on commit messages. Each pre-commit source code is used as the buggy code and each post-commit source code is used as the fixed code. They normalized the names of variables and custom methods to limit vocabulary size. Moreover, they removed samples including lexical or syntactical errors to limit the dataset to logical errors. They also filtered out the samples with more than 100 atomic AST modifications. This is a limit on their dataset that does not include bug-fix pairs requiring more modifications.

\citeauthor{lu2021codexglue} used the explained Java dataset to train and test their proposed CodeXGLUE APR method. They divided the dataset into two subsets named ``Small'' and ``Medium'' based on code lengths. The Small dataset includes samples with less than 50 tokens. The Medium dataset includes the samples with more than or equal to 50 and less than 100 tokens. They showed that CodeXGLUE is more effective on the Small dataset. We replicate all experiments in both datasets to extend the discussions of our findings.

We use CodeXGLUE to extend our study to a different APR model and further discuss a generalization of our findings. CodeXGLUE is based on transformers \cite{vaswani2017attention} that are the trend of machine learning and natural language research. CodeXGLUE leverages CodeBERT, a successful programming language model. This is an essential distinction between CodeXGLUE and TFix. TFix relies on NLP information and uses no software language model. Moreover, CodeXGLUE uses a single declaration line for each function. It uses no metadata or context. On the contrary, TFix uses surrounding code lines (context) and the static analyzer's metadata. Furthermore, CodeXGLUE is experimented on a Java dataset, while TFix is based on JavaScript. Java and JavaScripts are fundamentally different languages. Java is a statistically typed OOP language, while JavaScript is a scripting language. Also, CodeXGLUE uses function declarations while TFix uses method bodies. CodeXGLUE uses AST and word normalization as the preprocessing. TFix uses pure text. Finally, CodeXGLUE has not been directly compared with TFix, to the best of our knowledge.

Therefore, we find CodeXGLUE an excellent method for our study. First, this method is on the cutting edge research of Automated Program Repair and uses novel techniques. Second, it is quite different than TFix. Third, it is tested on a fundamentally different dataset of Java functions.

\subsection{Domain Adaptation}
In recent years, deep learning models \cite{lecun2015deep} have caused impressive advances in various tasks, including Computer Vision and NLP. 
However, deep learning-based models perform less well when applied to out-of-domain data. For example, a model trained on ``political news'' loses its performance when tested on ``sports news''. This issue is known as domain shift. Domain shift harms the ability of generic pretrained language models such as BERT \cite{devlin2018bert} and GPT-3.

Definition of domain depends on the context of each problem. \citeauthor{koehn2017six} \cite{koehn2017six} state that domains differ in terms of ``topic, genre, style or level of formality''. \citeauthor{saunders2021domain} \cite{saunders2021domain} further refine this definition and state that different domains come from different sources. Domains might be similar and overlap with each other or might be fundamentally different. For example, Covid news and Flue news are similar, while English and French texts are separate domains. The domain of an input sentence might be known or unknown. For example, if an article is collected from a health journal, the domain is known as health. On the other hand, if the same article is collected from an email box, the domain is not explicitly known. Lastly, domains might be few and limited or infinite and growing. For example, the number of genres in a website is limited. But, if each website is assumed as a separate domain, the number of domains will be infinite.

The following section explains the important related DA methods in NLP.

\vspace{1mm}
\subsubsection{An Overview of Important Domain Adaptation Methods in NLP}
\label{curr_bg}
Since domain adaptation in software engineering is very new (See section \ref{sc:related_work} for the existing work in this domain) domain adaptation in program repair and in general in software engineering is mainly the adopted or modified version of Domain Adaptation Methods in NLP. Therefore, in this section, we briefly review the concepts from this area that are needed to understand the rest of the paper.

A baseline approach for domain adaptation is to add new data to the training set and re-train the model, whenever a new domain is discovered. However, this approach is very inefficient since the initial training is a heavy process requiring substantial time and resources. Besides, this might be impractical in many cases as the training data and scripts are not always accessible.

Some DA methods are introduced that take a pretrained model and adapt (fine tune) it on the target domain's data. \full{} was used for NMT by \citeauthor{luong2015stanford} \cite{luong2015stanford} for the first time as a straightforward and efficient DA method \cite{saunders2021domain}. They showed that accuracy can improve significantly for in-domain data with low computation requirements \cite{luong2015stanford}. However, variations of \full{} have been proposed by changing model architecture, training scheme, or data to achieve better ``performance'', and ``efficiency'' or ``reliability''. We will explain these evaluation metrics in \ref{da_evaluation_bg}.

Freezing the whole network and tuning only a part of the model is another method of domain adaptation. This method aims to improve the efficiency while preserving general information of the model. \citeauthor{li2016one} \cite{li2016one} and \citeauthor{bapna2019simple} \cite{bapna2019simple} are two examples. The former introduces a new component to each hidden unit in the model and the latter adds new layers, called ``adapters''. In each case, the domain-specific multiplicative unit or adapter layer is tuned on the target data while the rest of the model is frozen. Adapter layers have achieved particular popularity as they require very simple and lightweight model modification. Also, they inherently involve no ``forgetting'' of general information since the original model parameters are unchanged. (Forgetting refers to the condition that the adapted model is no longer able to perform well on the source data). Notably, \citeauthor{abdul2020limsi} \cite{abdul2020limsi} find that adapters can outperform \full{} when translating to a ``noisy'' domain. In this condition \full{} model overfits on the noisy data. 

Curriculum learning is the idea of feeding training samples in an easy to hard order. It lets the model gradually learn the target domain. Variations of curriculum learning have been leveraged in DA for NMT. In some studies, they lead to higher effectiveness \cite{van2017dynamic, wang2018denoising, zhao2020reinforced}. Curriculum learning might degrade the efficiency due to the extra computation cost of reordering the samples. However, it might also improve the efficiency as the model converges faster. \cite{saunders2021domain}. Curriculum learning is also important in terms of ``Exposure Bias''.

In the domain adaptation literature, there three main evaluation criteria: Effectiveness, Efficiency, and Exposure Bias. Effectiveness refers to the accuracy of the adapted model in the target domain. Efficiency is about the time and memory requirement for the adaptation. 
Exposure bias refers to the situation when the model achieves excellent results on the adaptation set while it performs poorly on relatively different data. This cast doubts on the model's reliability and the results. This also shows the model is fragile to the upcoming changes and requires frequent retraining \cite{farajian2017multi, wang2020exposure, saunders2020addressing}. Exposure bias usually happens when the adaptation dataset is too small. Some studies show curriculum learning is more reliable in terms of exposure bias. The reason is that it smoothly transits to the target domain. This prevents sudden changes in the model and prevents it from overfitting on the target data \cite{chu2017empirical, aljundi2019gradient, song2019code}.

The explained domain adaptation methods require the presence of sufficient labeled data, known as parallel data. However, the majority of machine translation datasets may contain little or no in-domain labeled data because target domains are usually small. In some conditions, partial samples are only available. They miss either inputs or labels. Various methods including zero-shot learning \cite{ji2020cross, philip2020monolingual, zhang2020improving}, data selection \cite{poncelas2019selecting, chen2020character, bapna2019non} and data synthesis \cite{kim2021using,li2020metamt, mehta2020simplify, schwenk2008investigations, lambert2011investigations, edunov2018understanding, jin2020simple, currey2020distilling} have been proposed to solve APR without labeled data. Data synthesis is the category that we use in this study. Data synthesis techniques artificially create labeled data that can replace the missing real data in DA methods.

\section{Domain Shift and Domain Adaptation in Automated Program Repair}
\label{sc:method}
In general, domain shift can occur in software engineering tasks in several ways. Three most commons of them are as follows: 

\begin{itemize}
\item \textbf{Language Change}: A model that is trained on languages with high resources, such as Java or JavaScript, is supposed to work on languages with low resources, such as SQL or R.
\item \textbf{Synthetic Data to Real Data}: A model that is trained on synthetic data is supposed to work on real data.
\item \textbf{Cross Project}: A model that is trained on one or several projects is supposed to work on a different project.
\end{itemize}

Most software engineering models are trained on data that comes from several various projects. This means the data is naturally divided into many domains (projects). However, the number of projects used in the training set (``source projects'') is limited compared to the unlimited number of projects in the world. Therefore, it is quite likely that the trained models are applied on new and different projects known as ``target projects''. The shift of data distribution and characteristics caused by the change between source and target projects is the ``Cross Project'' domain shift we research in this study. In the following, we express some more characteristics of this type of domain shift (``Cross Project''):

\begin{itemize}
\item{} Our domains (projects) are normally known at both training and inference time. Each data sample belongs to a specific project. 

\item{} Data points in one project (domain) are not fundamentally different from data points of other projects. They use the same language, such as Java or JavaScript. They might even overlap; for example, two samples of different projects might look similar. The difference between domains is more in the distribution of data points. The coding styles or error types might differ from one project to another. Some domains are similar, and some are different. Consequently, the model may perform well on a new domain without any modification, but there is still the potential damage of domain shift and loss of effectiveness on other domains.

\item{} Efficiency can be a critical factor in APR tools. The number of domains might increase rapidly in software engineering. For a single company or workspace, many projects or sub-projects usually exist. 

\item{} 
The APR models might need to produce live and real-time recommendations while developers are changing source code, where a short inference time becomes important. If APR models are integrated into automated deployment, the time needed to analyze and fix thousands of code lines must not add high delays to CI/CD pipelines. In such scenarios, there is an essential need for efficient adapted models.

\item{} 
Collecting APR labeled data is not a trivial task. For example, only eight projects have more than 150 samples in the dataset of TFix \cite{berabi2021tfix}, and only 11 projects have more than 50 samples in both datasets of CodeXGLUE \cite{lu2021codexglue}. It normally requires strong filters that remove a high portion of collected data. 

\item{} On the other hand, monolingual (unlabeled/partial) APR data is always provided. One can assume the existence of substantial ``correct'' lines of code for any project. These codes act as unlabeled datasets. This provides an opportunity for a semi-supervised APR by creating synthetic buggy code out of the existing correct lines of code. 

\end{itemize}

In the current literature of APR studies, pretrained models are usually provided for the downstream projects. The users are supposed to pick them up as-is and use them for their own source code. This is the default approach that has no overhead or extra cost, but unfortunately suffers from the potential issues of domain shift. 

An alternative approach is to add the target projects' data to the training set. We use this approach as a baseline in this study, which is expected to avoid domain shift. However, the baseline approach is inefficient. It needs the full training of the model which requires huge resources such as time. Sometimes it is not even feasible since source datasets or training scripts are not accessible. Moreover, one can expect DA methods to show more effectiveness than full retraining, given that they are fine-tuned for each particular target project.

In the following two sections, we first propose a domain adaptation framework with three supervised methods. Then, we propose a data synthesis method and explain how it can complement the DA framework. Lastly, we indicate the potential changes needed to transfer the idea and methods of this work to other APR techniques.

\subsection{Domain Adaptation Framework for APR models}
\begin{figure}
  \centering
  \includegraphics[width=\linewidth]{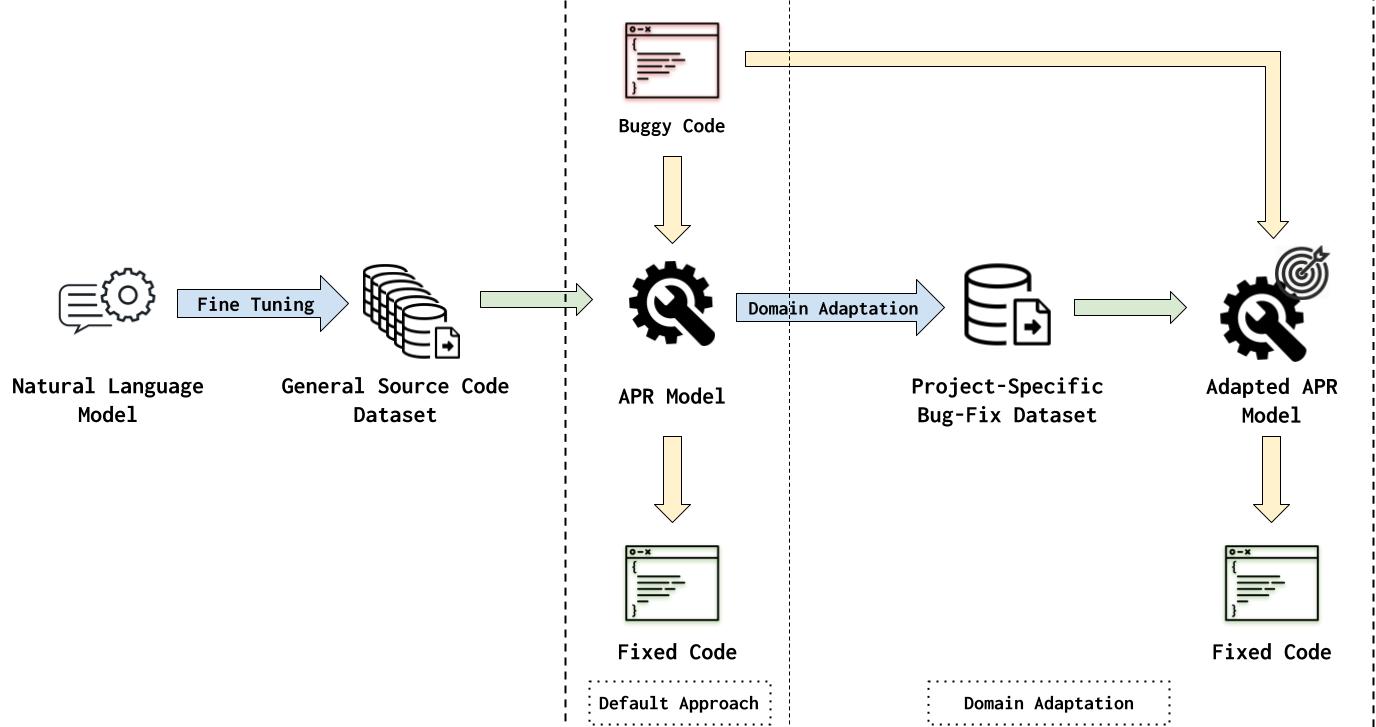}
  \caption{The Process of the Default APR Approach vs. the Domain Adaptation Framework for APR.}
  \label{fig:domain_adaptation_framework}
\end{figure}

The domain adaptation framework we propose is supervised. We assume access to the labeled data of the target project. The overall framework is depicted in Figure \ref{fig:domain_adaptation_framework}. The pretrained APR model is taken as-is without prior modifications in the training scheme or training data. Afterward, it is adapted to the target project using one of the proposed domain adaptation methods. Finally, the adapted model creates fixes for the bugs in the test samples of the target project. In the following, we describe three proposed domain adaptation methods we explored: ``Full Fine Tuning'', ``Tuning With Lightweight Adapter Layer'', and ``Curriculum Learning''.

\vspace{1mm}
\subsubsection{Full Fine Tuning}
We define ``Full Fine Tuning'' as continuing the optimization of the whole pretrained network on the target project. We take the pretrained model, feed it with the target project's data, in a random order, and update the weights using backpropagation. We repeat this cycle until the model converges. This is a straightforward and efficient method compared to the baseline approach. The model is optimized only for the small dataset of the target project instead of the whole corpus. 

\vspace{1mm}
\subsubsection{Tuning with Lightweight Adapter Layers}
\begin{figure}
  \centering
  \includegraphics[width=\linewidth]{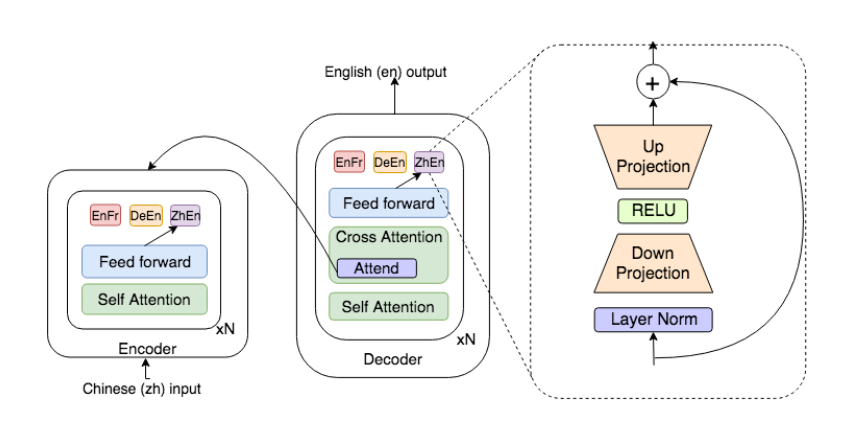}
  \caption{Overview of Tuning with Lightweight Adapter Layers \cite{bapna2019simple}.}
  \label{fig:adapter}
\end{figure}

We implement the method of \citeauthor{bapna2019simple} \cite{bapna2019simple} and call it ``Tuning With Lightweight Adapter Layers''. We take the pretrained model and freeze all its parameters to preserve the information on general data. Then, we incorporate ``lightweight adapter layers'' for domain adaptation. These layers are inserted between every layer of both the encoder and the decoder as shown in Figure \ref{fig:adapter}, within a dashed-border box. These layers are randomly initialized and will be optimized during domain adaptation.

Light-weight adapter layers are tiny modules with a tiny size compared to the whole network. They consist of a single hidden-layer feed-forward network with a non-linear ReLU\cite{agarap2018deep} activation function. The dimension between two layers is a hyper-parameter to be tuned. A normalization layer before the input is added to make the module pluggable to any transformer layer. A no-op residual module is also added that can bypass to let the model skip the projection.


\vspace{1mm}
\subsubsection{Curriculum Learning} 
Curriculum Learning is basically ``Full Fine Tuning'' with two important differences. First, the order of training samples is not random. Second, the amount of training data is increased in each epoch. We are inspired by the methods described in \ref{curr_bg} and implement our own method of Curriculum Learning based on APR's assumptions. These assumptions are: (a) we do not change the pretrained model training scheme or data, and (b) we do not directly use data from the general corpus in the adaptation. This is consistent with the assumptions of other DA methods. That is we take the pretrained model and only change the adaptation scheme.

We introduce a curriculum that ranks the training samples based on difficulty or distance. In each epoch, the model trains on the easier samples first and then learns the harder ones. We also start with a small portion of training samples and increase it in each epoch. The idea is to gradually adapt the model to the target data, also known as ``soft adaptation''. The rate of the samples fed to the model is as Table \ref{tab:curr_portion}.

\begin{table}
  \caption{Tuning Data Portion in each Epoch of Curriculum Learning}
  \label{tab:curr_portion}
  \centering
  \begin{tabular}{lccc}
    Epoch&1&2&3, etc\\
    \midrule
    Portion&35\%&70\%&100\%\\
\end{tabular}
\end{table}

The result of Curriculum Learning hugely depends on the type of the used curriculum. We propose the following three types of curriculum.

\vspace{1mm}
\begin{itemize}
\item \textbf{Code Length}: Longer code lines are harder to fix for the model \cite{lu2021codexglue}.
\item \textbf{Model Confidence}: The model returns a confidence score for each prediction. The lower the score, the harder the code line is.
\item \textbf{Similarity to General Data}: More similar a data point is to the general data, the easier it is to fix.

We note that the last curriculum requires access to the source (general) dataset but does not change the training scheme of the pretrained model. We use CodeBERT\cite{feng2020codebert} as the tool to extract embeddings from samples. Then, we measure the cosine similarity \cite{singhal2001modern} between the vectors of the source dataset and each sample of the target project.
\end{itemize}

\subsection{APR Data Synthesis} 
\begin{figure}[h]
  \centering
  \includegraphics[width=\linewidth]{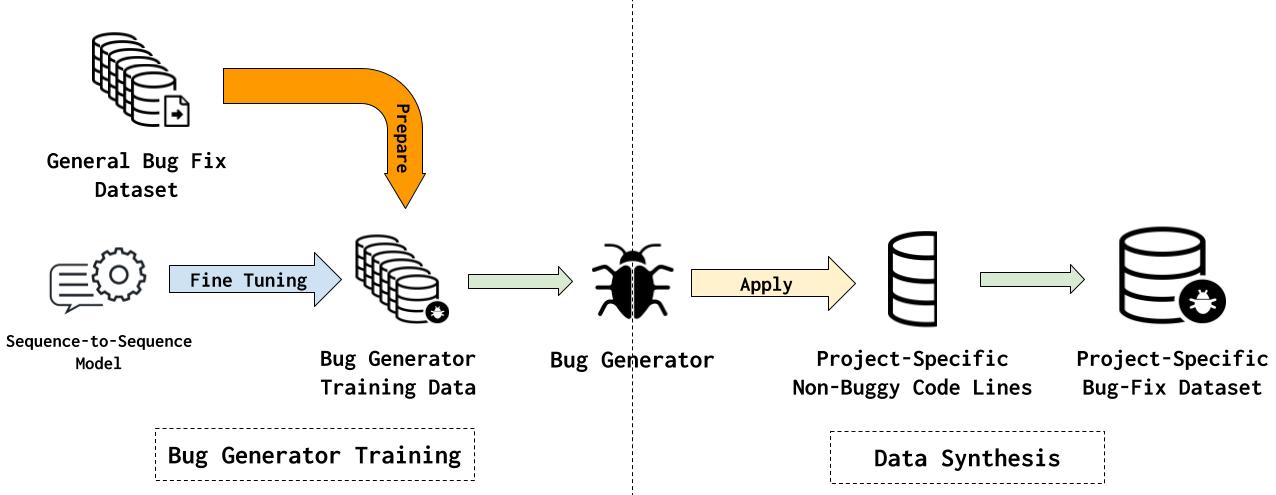}
  \caption{Overview of APR Data Synthesis Method.}
  \label{fig:proposed_data_augmentation}
\end{figure}

In this section, we explain our proposal for an APR data synthesis method. The goal is to create (extra) labelled data (i.e., pairs of buggy code and its fixed version) for project-based domain adaptation. Note that the synthetic data is only used for fine-tuning and the final evaluations are based on real bugs and generated fixes. 
Our overall idea (depicted in Figure \ref{fig:proposed_data_augmentation}) is to create bugs with a ``reverse'' model corresponding to the APR model under study. This mean the bug generator model creates buggy code from clean code. In the rest of this section, we explain our bug generation process and model and show how it integrates into the domain adaptation framework.

\begin{figure}[h]
  \centering
  \includegraphics[width=0.75\textwidth]{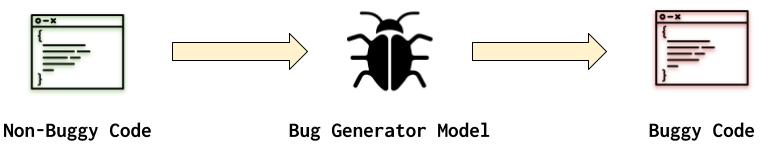}
  \caption{Bug Generator Model.}
  \label{fig:tbug_inference}
\end{figure}

The bug generator model (as depicted in Figure \ref{fig:tbug_inference}), gets non-buggy code and injects a bug in it. Similar to APR models, we formulate the task of bug injection as a machine translation task. Note that although there exists classic bug generation techniques such as mutation generation, however, we opt for this synthesizing approach given that it can create more similar bugs to the real bugs from the APR dataset. It is interesting to compare our bug generator approach with a mutation generation tool, but that goes beyond the scope of this paper (studying DA methods for APR).

Figure \ref{fig:proposed_data_augmentation} shows how we train the bug generator model, and, how we use it to synthesize APR dataset for a specific project. First, the training data is prepared mainly by swapping the input and output of a bug-fix dataset. That means the inputs are non-buggy code lines, and the labels are buggy code lines. Then, a sequence-to-sequence model is fine-tuned on this dataset. This is similar to the training process of APR models. The model learns how to get a non-buggy code snippet and converts it to a buggy code snippet, during training epochs.

Afterward, we apply the trained bug generator model to a target dataset. Each code line and the generated bug create a bug-fix instance. These instances together create a project-specific bug-fix APR dataset for the target project, to be used in the DA framework.

Our method still assumes the presence of a general bug-fix dataset. This is a reasonable assumption. General APR datasets are provided for various languages \cite{lu2021codexglue, tufano2018empirical, berabi2021tfix}. The main benefit of this method is to create \textbf{project-specific} bug-fix dataset, which is usually lacking in practice for the project in hand.

The choice of the Sequence-to-Sequence model can be based on the APR model. For example, we use T5 for TFix and CodeBERT-Decoder for CodeXGLUE. There are mainly two reasons for that. First, each architecture is showed to be effective for the corresponding dataset. For example, CodeBERT-based Encoder-Decoder is effective for the dataset of CodeXGLUE\cite{lu2021codexglue}. Therefore, we expect it to be effective for bug generation on the same type of dataset (both datasets have similar structures. For example, they are both pure JS codes and do not use AST of a Java code that has some specific postprocessing). Second, the bug generator model will have the exact requirements as the APR model, which makes that more practical. That is, the system that can run the APR model most likely will run the bug generator model smoothly.

Furthermore, we note that injecting bugs is different from fixing them in the way that there are more diverse ways to create bugs, but fixing a bug can be assumed more unique. For a non-buggy line, there are many types of bugs to create. This prevents the model from learning the bug generation during training. Whenever the model creates correct buggy code that is not identical to the buggy line, it gives misguiding network updates. To alleviate this issue and improve the effectiveness of the bug generator models, we augment the input data with the \textbf{bug types}. Therefore, the format of the bug generator will be as follows:

\vspace{2mm}
\begin{center} bug \fcolorbox{pink}{white}{error type} \textcolor{teal}{non-buggy line} $\Longrightarrow$ \textcolor{red}{buggy line} \end{center}
\vspace{2mm}

Using bug-type metadata, the model knows which type of bug it is expected to generate. This reduces the number of potential outputs (bugs) for each input (non-buggy code line) which guides the model more accurately. The model learns how to generate different types of bugs (based on the bug-type metadata) through training epochs.

Note that adding bug types metadata to the inputs of the bug generator model is not always practical. For example, CodeXGLUE does not come with bug type meta data. One can potentially use static analyzer tools for Java such as LightRun \cite{lightrun} and add bug types to the dataset. However, we do not have any evidence of that with CodeXGLUE dataset. Therefore, we only study it in the context of TFix.

Also note that the other optimizations and features that are implemented by APR models can be leveraged into the correspondent bug generator. For example, a technique might use special type of preprocessing (such as AST generation) for a dataset.  To summarize, we conclude that the choice of bug generation pipeline, including the sequence-to-sequence model, the bug type metadata, and other optimizations can be determined by the pipeline of the corresponding APR model.

\section{Experiments}
\label{sc:experiments}
\subsection{Objectives and RQs}

The objective of the experiments is to analyze the effect of domain shift and domain adaptation on TFix and CodeXGLUE. The following is the list of research question with a short motivation per RQ. We will describe the design and result per RQ in Section \ref{sc:evaluation}.

\begin{itemize}
\item \textbf{RQ1: \rqa }
In this RQ, as a fundamental motivation of this study, we design an experiment to measure the potential damage of domain shift to TFix..
\vspace{2mm}

\item \textbf{RQ2: \rqb }
\begin{itemize}
    \item \textbf{RQ2.1: \rqbw }
    \item \textbf{RQ2.2: \rqbx }
    \item \textbf{RQ2.3: \rqby }
\end{itemize}

\vspace{1mm}
In RQ2, we employ our proposed domain adaptation framework and analyze its improvements to TFix compared with the default and baseline approach. We thoroughly study the three DA methods and compare effectiveness, efficiency, and reliability (exposure bias) in RQ2.1, RQ2.2, and RQ2.3, respectively.
\vspace{2mm}

\item \textbf{RQ3: \rqc }
\begin{itemize}
    \item \textbf{RQ3.1: \rqcw }
    \item \textbf{RQ3.2: \rqcx }
\end{itemize}

\vspace{1mm}
In RQ3, we test the proposed data synthesis method in case of no data. We analyze the accuracy of TBug itself in RQ3.1 and its improvement to the DA framework in RQ3.2.
\vspace{2mm}

\item \textbf{RQ4: \rqd }
\begin{itemize}
    \item \textbf{RQ4.1: \rqdw }
    \item \textbf{RQ4.2: \rqdx }
    \item \textbf{RQ4.3: \rqdy }
\end{itemize}
\vspace{1mm}
Finally, we replicate the key findings to CodeXGLUE in RQ4. In RQ4.1, we measure the effect of domain shift on CodeXGLue. In RQ4.2, we measure the improvement of the DA framework with \full{} on CodeXGLUE. In RQ4.3, we measure the improvement of the data synthesis method using CodeXBUG. 

\end{itemize}

\subsection{Datasets and Models}
\subsubsection{Datasets}
In this study, we use two datasets. JavaScript (TFix) and Java (CodeXGLUE).

\begin{itemize}
\item \textbf{RQ[1-3]:} We use the JS dataset created by \citeauthor{berabi2021tfix} and only change the train, validation, and test splits according to our design. We do not change or remove samples. We include all projects and error types.

\item \textbf{RQ4}: We use the dataset of Java functions created by \citeauthor{tufano2019empirical}, which was used in the original study of CodeXGLUE \cite{lu2021codexglue}. We replicate all experiments in both Small and Medium dataset types to add to the generalizability of our discussions. We only change train, validation, and test splits according to our design. We include all the projects and do not change or remove any samples. 
\end{itemize}

We label each input sample in all datasets with the project from which it is extracted. This label is already included in the dataset of TFix. For CodeXGLUE, we pulled it from the parent datasets.

\vspace{1mm}
\subsubsection{Models}
\paragraph{TFix}
\citeauthor{berabi2021tfix} used five configurations of T5 that are T5-Small, T5-Large, T5-Base, T5-Large-no-pretrain, T5-Large-per-type.  They showed T5-Large is the most effective configuration. In this work, we use T5-Large as the main configuration to study and improve. In addition, we note that T5-Small has advantages in terms of efficiency. It is lighter and needs fewer resources. We also expect T5-Small to behave differently in domain shift. Smaller and larger models are differently affected when the distribution of test data changes. Therefore, we recognize T5-Small as a representative of a small APR model and fully expand our experiments to it. We refer to the resulting models, TFix-Large and TFix-Small, respectively. In the rest of this study, we refer to those configs as TFix-Large and TFix-Small. We also use the same architectures for TBug and call them TBug-Small and TBug-Large.

\paragraph{CodeXGLUE} 
CodeXGLUE comes with a single Encoder-Decoder architecture. CodeBERT is used as the encoder, and the decoder is initialized randomly. We use this model architecture without any modification for both CodeXGLUE and CodeXBUG.

\vspace{1mm}
\subsubsection{An Experiment Note} 
We note that CodeXGLUE is weaker than TFix. APR is not the central focus of CodeXGLUE paper. It is not compared to any novel APR methods rather than a naive baseline. The dataset is smaller and more limited than the one of TFix. For example, if we filter the projects that have at least 150 samples, the resulting set includes 24 projects for TFix. This number is only one project for CodeXGLUE (in each dataset). Therefore, we conduct the full empirical study with both sizes of TFix. But, limit the experiments to the key findings for CodeXGLUE in RQ4.

\subsection{Evaluation Metrics}
In this section, we explain the evaluation metrics used in the experiments.

\paragraph{Effectiveness} 
Effectiveness is used in RQ1, RQ2.1, RQ3, and RQ4.1. It is the primary evaluation metric in APR studies. It measures how well the model predicts the desired changes of developers. Most well-known APR studies use \textbf{``Exact Match''} as the effectiveness metric. Exact Match refers to the ratio of generated fixes identical to the changes made by developers. Although a correct fix is not necessarily unique and an APR may create a correct patch that is not an exact match, given that most studies use ``exact matc'', we also use it as our effectiveness metric.

BLEU score \cite{papineni2002bleu} is another widely used metric in NLP, Software Engineering, and APR. It is known to be correlated with the quality of generated code/texts \cite{papineni2002bleu}. However, some studies show that the BLEU score is neither necessary nor sufficient for achieving an actual improvement \cite{callison2006re}. CodeBLEU score \cite{ren2020codebleu} is a similar metric to BLEU score, specialized for programming datasets. Similarly, this score is not as accurate as Exact Match \cite{callison2006re}. Also, neither BLEU nor CodeBLEU scores are used by \citeauthor{berabi2021tfix} in TFix's original paper. Thus, we do not include them in this study as well.

\citeauthor{berabi2021tfix}, instead, introduced a novel metric called ``Error Removal''. Error removal refers to the number of errors that disappear from error detector results. However, this metric requires a complex algorithm. For example, if one error is removed and three other errors are introduced due to a modification, it is unclear if the code is improved or impaired. \citeauthor{berabi2021tfix} are the only ones who used this metric to the best of our knowledge. They did not publish their implementation or exact algorithm. Therefore, we do not include error removal in this study.

\paragraph{Efficiency}
We  measure efficiency of DA methods in RQ2.2, by the following metrics:

\begin{itemize}
\item \textbf{Model Preparation Time}: This is the time taken by the system to prepare the ready-to-use models. This is a one-time process but might be repeated every time that data changes significantly. For the DA framework, the preparation time is the time for fine-tuning the model on the target project.

\item \textbf{Inference Time}: Inference time is the average time to fix a bug.

\item \textbf{Model Size}: Model sizes are measured by the storage volume they use in the memory. Their volume highly correlates to the run-time storage they reserve such as RAM or GPU memory. 
\end{itemize}

All efficiency metrics depend on the configurations and the resources of systems. We measured them mainly for comparisons. Moreover, we run all experiments on the same system with similar configurations such as batch size, saving method, and early stopping strategy. This makes the comparisons sound.

\paragraph{Exposure Bias}
Exposure bias is measured in RQ2.3. It is similar to effectiveness. The only difference is the dataset under the experiment. Exposure bias is calculated on different data points than the target project (we use samples of general data), while effectiveness is measured on the target project's data. Therefore, we use the same ``exact match'' metric for exposure bias.

\subsection{Experiment Setup}
We use the scripts published by authors of TFix and CodeXGLUE to run the experiments. The only change we make in creating the pretrained model is the data division, which will be further explained in \ref{sc:evaluation}. We do not change other configurations and hyperparameters.

We use a single node of ComputeCanada (Cedar) for all experiments with 32GB v100l GPU, 4 CPU cores, and 30GBs RAM. All experiments are possible to do with 16GB GPU, 1 CPU core, and 20 GB RAM with minor changes in the timings. These are not strict requirements supporting the practicality of the methods in personal systems.

\section{Evaluation}
\label{sc:evaluation}
This section explains each research question's design and results.

\subsection{RQ1 Evaluation) \rqa{}}

\subsubsection{Design}
We select projects with at least 150 samples as the potential target projects. We argue that the projects with very few samples are not proper candidates for the evaluation purposes. Since the domain shift's impact will be major if the target project data is small and very few samples from the project are in the training data. Moreover, in the next RQs, the proposed DA methods are not expected to perform well with very few adaptation samples. The test results will also be unreliable if the number of the test samples is less than 30. 

Using the explained filter, the total number of selected projects is 24. We sort them based on their sizes and select every third of them as  ``Target Set'' and leave the rest as ``Source Set''. The source set plays the role of general data on which the model is trained. The target set represents a collection of target projects (new projects) on which the model is tested. The statistics of projects in each set are reported in Table ~\ref{tab:stats}.

\begin{table}
  \caption{Statistics of Source and Target Sets.}
  \label{tab:stats}
  \centering
  \begin{tabular}{ccc}
    \midrule
    Set&Projects&Samples\\
    \midrule
    Source&23277&102908\\
    Target&8&1896\\
  \bottomrule
\end{tabular}
\end{table}

\citeauthor{berabi2021tfix} \cite{berabi2021tfix} report that the accuracy of TFix varies between different error types. Thus, they split the data into train, validation, and test splits per each error type. We follow the same design. We also want the data of each project to be divided proportionally between the splits. Therefore, we split the data as follows. For each pair of a project and an error type, we split the corresponding data points into train and test splits with the ratio of 80\%, and 20\%. Then, we aggregate the data points of all error types in each project. We also further divide train splits into train and validation. In the end, the target data is divided according to Table \ref{tab:projects_statistics}.

There is a note in splits preparation. The number of test samples of each error type is up-rounded. For example, if there are only two samples in a project, one of them is assigned to the training split and one to the test split. Therefore, there is a difference between the actual number of test samples in each project and the expected 20\%.

In the rest of the experiments, we call the aggregated samples in each set ``source-train'', ``source-test'', ``target-train'', and ``target-test''. This is important to finalize the splits of all samples at this step, since we want to keep the test set the same between RQs[1-3]. Although ``source-test'' and ``target-train'' are not used in this research question, they are prepared for the following research questions. 

\begin{table}[]
\caption{Statistics of Target Projects.}
\label{tab:projects_statistics}
\centering
\begin{tabular}{lccccc}
\toprule
\#              & project      & \# Samples & \# Train & \# Validation & \# Test \\
\midrule
1                    & Qooxdoo      & 547        & 345      & 87            & 115     \\
2                    & Kibana       & 231        & 144      & 36            & 51      \\
3                    & Ember.js     & 218        & 136      & 35            & 47      \\
4                    & Core-js      & 201        & 128      & 32            & 41      \\
5                    & ONM          & 194        & 124      & 31            & 39      \\
6                    & Sequelize    & 180        & 112      & 28            & 40      \\
7                    & Dcos-ui      & 166        & 103      & 26            & 37      \\
8                    & LivelyKernel & 159        & 98       & 25            & 36      \\
\midrule
\multicolumn{1}{l}{} & Sum          & 1896       & 1190     & 300           & 406   \\
\bottomrule
\end{tabular}
\end{table}

\begin{figure}[h]
  \centering
  \includegraphics[width=\linewidth]{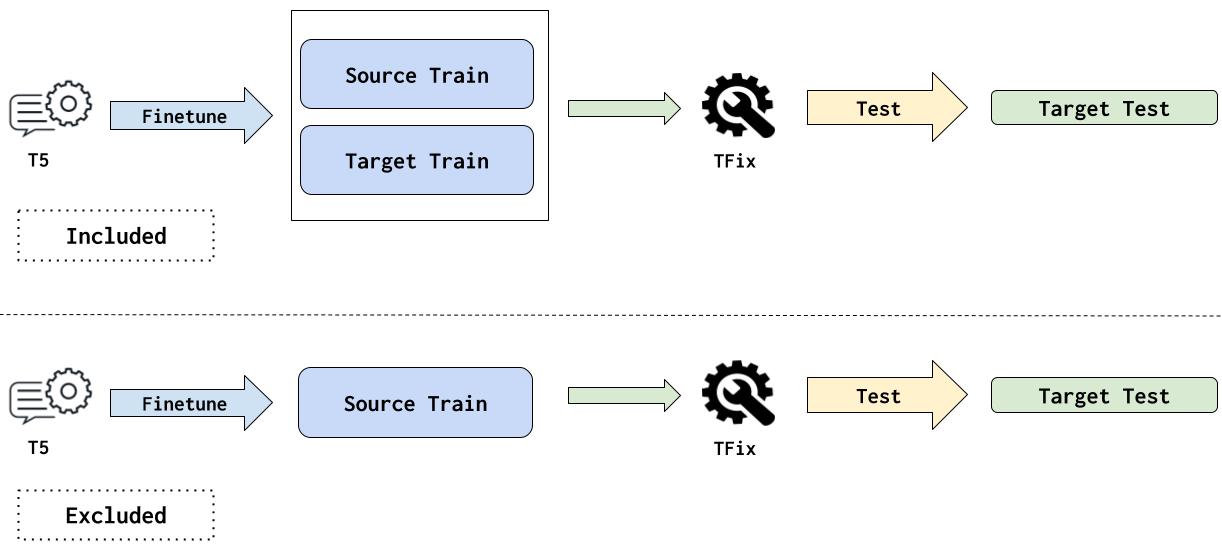}
  \caption{Included and Excluded Design Scenarios of TFix Training.}
  \label{fig:rq1}
\end{figure}

We cannot use the pretrained model of TFix provided by \citeauthor{berabi2021tfix} because we have changed the train, validation, and test splits. We re-train TFix from scratch in two included and excluded scenarios. They are depicted in Figure \ref{fig:rq1} and explained below.

\begin{itemize}
\item \textbf{Included:} In the Included scenario, the model is trained jointly on source-train and target-train. The samples are all equally important to the model regardless of the set they belong to. This design assumes that the projects of the target set are accessible during the training and represents the case of ``without domain shift'', which is the implicit assumption of the original study of TFix. 

\item \textbf{Excluded}: In the excluded scenario, the model is trained only on source-train. The model does not get any data points from the target set during the training. In other words, it assumes the test projects are unavailable during the training. It represents the real-world case in which domain shift is expected. 
\end{itemize}

We take only 8 of 24 projects for the target set (less than 2\% of the whole dataset \ref{tab:stats}). The main reason is to make sure the excluded data do not cover a vast portion of the main dataset. Otherwise, the potential drop in the accuracy could have been mainly due to fewer training samples. 

Finally, we test both models on the test splits of each 8 target projects. In fact, instead of shifting the test data, we keep the test domain the same and shift the training data (by excluding the training samples of the target projects from the training set). This makes the results of Included and Excluded scenarios comparable as both are tested on the same test data. We repeat all design step explained here for both T5-small and T5-large architectures and compare them.

\subsubsection{Results}

\begin{table}[]
\caption{Project-wise Exact Match Accuracy of the Included and Excluded Experiments on Target Projects with TFix-Small and TFix-Large.}
\label{tab:comparison_excluded}
\centering
\begin{tabular}{cllcccc}
\multicolumn{3}{c}{}&\multicolumn{2}{|c|}{TFix-Small}&\multicolumn{2}{|c}{TFix-Large} \\
\toprule
  & Project       & Samples       & Included       & Excluded      & Included       & Excluded      \\
\toprule
1                     & Qooxdoo       & 547              & \textbf{53.04}          & 48.70         & \textbf{58.26}          & 46.08         \\
2                     & Kibana        & 231              & \textbf{45.10}          & 43.14         & \textbf{60.78}          & 45.10         \\
3                     & Ember.js      & 218              & \textbf{53.19}          & 51.06         & \textbf{65.96}          & 51.06         \\
4                     & Core-js       & 201              & \textbf{24.39}          & 21.95         & \textbf{51.22}          & 39.02         \\
5                     & ONM           & 194              & 100            & 100           & 100.00          & 100.00         \\
6                     & Sequelize     & 180              & 35.00          & \textbf{40.00}         & \textbf{55.00}            & 40.00           \\
7                     & Dcos-ui       & 166              & 94.60          & 94.60         & \textbf{86.49}          & 67.57         \\
8                     & LivelyKernel  & 159              & 61.11          & 61.11         & \textbf{69.44}          & 66.67         \\
\midrule
& \multicolumn{2}{c}{Weighted Average}  & \textbf{56.40}          & 54.93         & \textbf{66.01}          & 54.19         \\
&                  \multicolumn{2}{c}{Average}          & \textbf{58.30}          & 57.77         & \textbf{68.39}          & 56.94         \\
&                \multicolumn{2}{c}{Median}           & \textbf{53.12}          & 49.88         & \textbf{63.37}          & 48.58  \\  
\bottomrule
\end{tabular}
\end{table}

Table \ref{tab:comparison_excluded} shows the exact match accuracy of TFix in the included and excluded experiments on target-test. The results are for both TFix-Small and TFix-Large. Each row corresponds to one of the 8 selected projects. Each cell is the exact match accuracy between generated fixes and the references. The higher number between the two designs is marked for each row and size. 

Firstly, statistics of projects show that we cover a diverse range of projects in this experiment. The number of samples varies from 547 to 159. The set includes easy and hard projects. For example, ``ONM'' and ``Dcos-ui'' have the accuracy of 100\% and 67.57\% while ``Core-js'' has the accuracy of 30.02\% (using TFix-Large in the excluded design).

Secondly, TFix-Large, which is the primary and more powerful model of TFix, loses its accuracy significantly. The drop is 14.79\% in Median and 11.82\% in the weighted average (more than 10\% in 6 projects, 2.77\% in 1 project, and 0\% in the easy ONM). However, TFix-Small's loses are lighter. The Median loss is 3.24\% and the weighted average of losses over projects is 1.47\% (the accuracy drops less than 5\% in 4, remains the same in 3, and increases in 1 of the projects). It is worth mentioning that with TFix-Small, all 4 projects that do not suffer from domain shift have less than 200 samples.

TFix-Large's significant effectiveness loss is most likely to do with its large capacity, many layers, and learnable parameters. They provide the model with the ability to overfit to the internal distributions of projects. The model learns how to perform differently for each specific target project according to its characteristics. However, when the access to the target-train data is removed, the model can no longer perform as well. In contrast, the plausible reason for TFix-Small's resistance toward domain shift is that the small model has fewer parameters that only learn the general patterns of bug fixes. This model does not discover particular patterns specific to each project. 

While TFix-Large has outstanding performance in the included design and is suggested as the best architecture in the original TFix paper, all of its advantages are gone in the more realistic scenario (Excluded). This shows that domain shift is a valid concern in the case of TFix. This further motivates us to employ domain adaptation methods to fix this deficiency and enable TFix-Large to benefit from its capacity.

\vspace{5mm}
\begin{mdframed}[style=mpdframe,frametitle=RQ1 Summary]
TFix-Large and TFix-Small show different levels of vulnerability to domain shift. While TFix-Small is relatively resistant to domain shift and its drop is negligible, TFix-Large hugely loses its effectiveness. This confirms the threat of domain shift to TFix and motivates the employment of domain adaptation to address the issue.
\end{mdframed}

\subsection{RQ2 Evaluation) \rqb}

\subsubsection{Design}
In this RQ, we conduct an experiment on three domain adaptation methods that fit into our proposed framework. We also include the default and baseline approaches. The default approach (Default) is to take the pretrained model, trained on the source set, and test it as-is on each target project. This approach is identical to the ``excluded'' design of RQ1. The baseline approach (Baseline) is to include the data of target projects in the training set, train TFix on them and test it on each project of the target set. This is equivalent to the ``included'' design of RQ1. Therefore, we borrow the results of included and excluded designs in RQ1 and use them as Default and Baseline.

We assume that the pretrained TFix model is provided. This model is published by \citeauthor{berabi2021tfix}. In our experiment design, we use the excluded model of RQ1 as the pretrained model.

We have three domain adaptation methods that are \full{} (\fullm), \adapter{} (\adapterm{}), and \curr{} (\currm{}). We use three types of \curr{} that are \currc{} (\currcm{}), \currl{} (\currlm{}), and \currs{} (\currsm{}). This creates 5 method types in total. Moreover, we have 8 test projects. Therefore, we run the DA framework for \ (5 * 8 =) 40 \ pair of DA methods and target projects.

\begin{figure}[h]
  \centering
  \includegraphics[width=\linewidth]{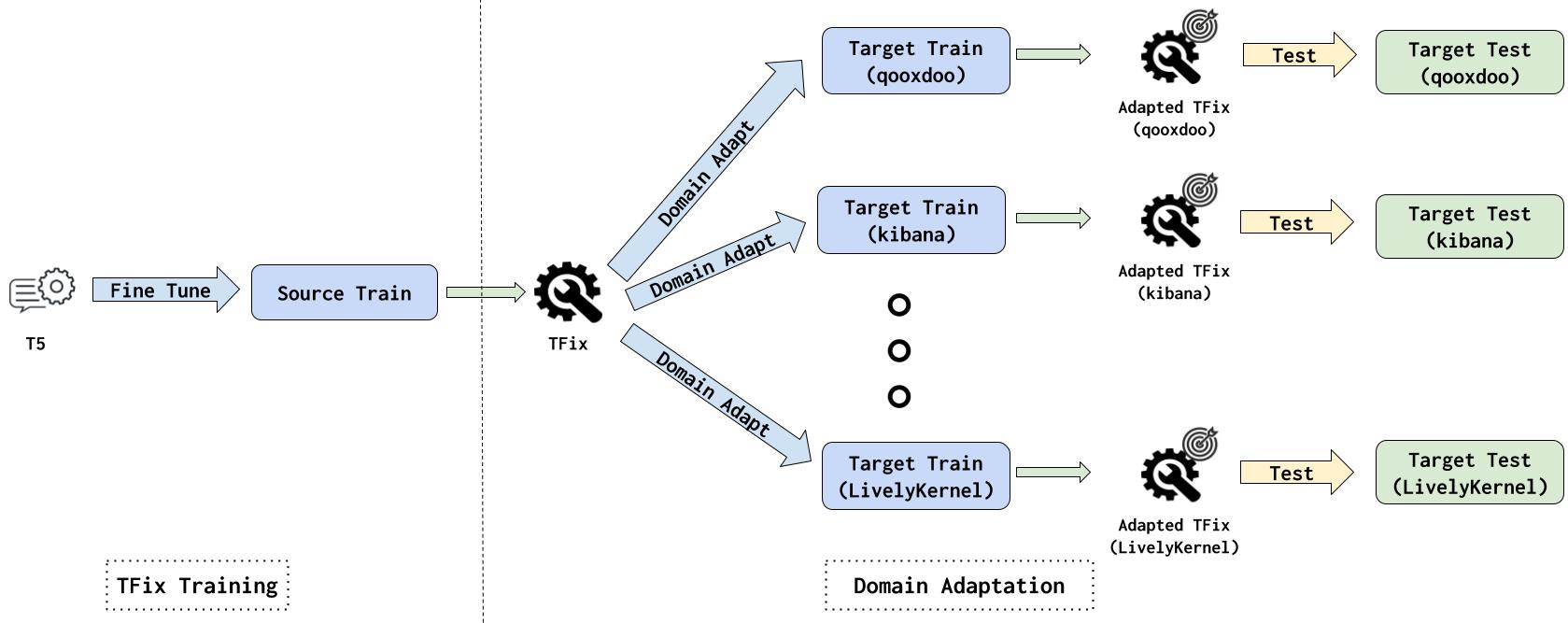}
  \caption{Experiment Design for Domain Adaptation of TFix to Each Target Project.}
  \label{fig:rq2}
\end{figure}

For each target project, we use the corresponding part of target-train for adaptation. Next, we test the resulting adapted model on the corresponding part of target-test. The explained experiment design is depicted in Figure \ref{fig:rq2}.

We compute the efficiency metrics explained in ``Evaluation Metrics'' section in RQ2.2. The \underline{preparation time} of the domain adaptation methods is the time taken to optimize the network on the target dataset. We domain adapt pretrained models on each of the 8 target projects and report the average time. The preparation time of the default approach is 0 because there is no extra process on the model. The training time of Baseline is equivalent to training the model with full data. Moreover, we feed the target-test data of each target project to the corresponding adapted model and measure the response time as the \underline{inference time}. Finally, we measure the disk space occupied by the save adapted models as the \underline{model size}.

We analyze the effect of exposure bias for each DA method in RQ2.3. The goal is to check if the model overfits the small target set. Therefore, we randomly take 5000 samples from source-test (general data). Then, we test the pretrained model and each of 40 adapted models on them. If the accuracy of adapted models is substantially lower than the pretrained model, they overfit the target data. In other words, they forget the general ability of program repair. However, if they keep the performance on general data, the models and results are reliable regarding exposure bias.

In all DA methods, we use EarlyStopping. It stops training epochs if no more improvement happens in four consecutive epochs. After each epoch, the model is evaluated, and the best version is stored. 

\subsubsection{Results}
In the following three sub-research questions, we analyze and compare the proposed domain adaptation methods in terms of Effectiveness (RQ2.1), Efficiency (RQ2.2), and Exposure Bias (RQ2.3).
 
\paragraph{RQ2.1) \rqbw}
Table \ref{tab:comparison_methods_large} and Table \ref{tab:comparison_methods_small} report the accuracy of Default, Baseline and each DA method. Each cell is the exact match accuracy between generated and actual fixes. The results are reported for each model size and project, separately. ``Weighted Average, ``Average'', and ``Median'' are the aggregated metrics used. We mark the highest value in each row and the numbers with less than 3\% differences. In the following, we analyze the results.

\begin{table}[]
\caption{Project-wise Exact Match Accuracy of Domain Adaptation Framework for TFix-Large}
\label{tab:comparison_methods_large}
\centering
\footnotesize
\begin{tabular}{ccccccccccc}
Project              & \# Tests         & Default & Baseline & \fullm{}           & \adapterm{}        & \currsm{} & \currlm{} & \currcm{} \\
\midrule
Qooxdoo                     & 115                  & 46.08   & 58.26    & 58.26          & \textbf{63.48}          & 59.13    & 48.70   & 47.83    \\
Kibana                     & 51                   & 45.10   & 60.78    & \textbf{66.67}          & 58.82          & 45.10    & 49.02   & 49.02    \\
Ember.js                    & 47                   & 51.06   & \textbf{65.96}    & 61.70          & 61.70          & \textbf{65.96}    & 53.19   & 53.19    \\
 Core-js                     & 41                   & 39.02   & \textbf{51.22}    & 46.34          & \textbf{51.22}          & 46.34    & 39.02   & 39.02    \\
ONM                         & 39                   & 100     & 100      & 100            & 100            & 100      & 100     & 100      \\
Sequelize                   & 40                   & 40.00   & 55.00    & \textbf{62.50}          & \textbf{60.00}          & 55.00    & 45.00   & 55.00    \\
Dcos-ui                     & 37                   & 67.57   & 86.49    & \textbf{94.59}          & 89.19          & \textbf{94.59}    & 78.38   & \textbf{97.30}    \\
LivelyKernel               & 36                   & 66.67   & 69.44    & 69.44          & 69.44          & 66.67    & 66.67   & 66.67    \\
\midrule
\multicolumn{2}{c}{Weighted Average} & 54.19   & \textbf{66.01}    & \textbf{67.24} & \textbf{67.49} & 64.29    & 57.14   & 59.61    \\
\multicolumn{2}{c}{Average} & 56.94   & \textbf{68.39}    & \textbf{69.94}          & \textbf{69.23}          & 66.60    & 60.00   & 63.50    \\
\multicolumn{2}{c}{Median} & 48.58   & \textbf{63.37}    & \textbf{64.59}          & \textbf{62.59}          & \textbf{62.55}    & 51.11   & 54.09   \\
\bottomrule
\end{tabular}
\end{table}

\begin{table}
\caption{Project-wise Exact Match Accuracy of Domain Adaptation Framework for TFix-Small}
\label{tab:comparison_methods_small}
\footnotesize
\centering
\begin{tabular}{ccccccccccc}
Project        & \# Tests        & Default & Baseline & \fullm{}           & \adapterm{} & \currsm{} & \currlm{} & \currcm \\
\midrule
Qooxdoo                   & 115                 & 48.70   & 53.04    & 52.17          & \textbf{58.26}   & \textbf{55.65}    & 49.57  & 49.57   \\
Kibana                    & 51                  & 43.14   & 45.10    & 49.02          & 47.06   & \textbf{52.94}    & 43.14  & 43.14   \\
Ember.js                  & 47                  & 51.06   & \textbf{53.19}    & \textbf{55.32}          & 51.06   & \textbf{55.32}    & 51.06  & 51.06   \\
Core-js                   & 41                  & 21.95   & 24.39    & \textbf{56.10}          & 46.34   & 46.34    & 21.95  & 46.34   \\
ONM                       & 39                  & 100     & 100      & 97.44          & 100     & 100      & 100    & 100     \\
 Sequelize                & 40                  & 40.00   & 35.00    & \textbf{50.00}          & 35.00   & 40.99    & 35.00  & 42.50   \\
Dcos-ui                   & 37                  & \textbf{94.60}   & \textbf{94.60}    & \textbf{97.44}          & 86.49   & 86.49    & \textbf{94.59}  & 86.49   \\
LivelyKernel              & 36                  & 61.11   & 61.11    & 61.11          & \textbf{66.67}   & 61.11    & 61.11  & 61.11   \\
\midrule
\multicolumn{2}{c}{Weighted Average} & 54.93   & 56.40    & \textbf{60.59} & \textbf{59.85}   & \textbf{60.34}    & 56.60  & 57.14   \\
\multicolumn{2}{c}{Average}          & 57.57   & 58.30    & \textbf{63.46}          & \textbf{61.36}   & \textbf{62.23}    & 57.05  & 60.03   \\
\multicolumn{2}{c}{Median}           & 49.88   & \textbf{53.11}    & \textbf{55.71}          & \textbf{54.66}   & \textbf{55.48}    & 50.31  & 50.31\\ 
\bottomrule
\end{tabular}
\end{table}

According to Table \ref{tab:comparison_methods_large} (T5-Large), \full{} and \adapter{} reach the highest effectiveness. Baseline is slightly weaker than them. \curr{} group have relatively lower effectiveness that varies from type to type. \currs{} is the most effective one that is still lower than the best methods. \currl{} and \currc{} are the weakest methods. All domain adaptation methods have a considerable improvement over Default. This improvement is huge for Baseline, \full{}, \adapter{}.

Table \ref{tab:comparison_methods_small} shows that Domain Adaptation has similar results in TFix-Small with few key differences. First, Baseline does not show the same level of effectiveness. It does not outperform Default with a notable margin and is much weaker than other DA methods. Second, \currs{} performs better in this size compared to the large size. Its effectiveness is similar to \full{} and \adapter{}. Third, the improvements caused by Domain Adaptation are less. Although the difference between Default and the best DA method is still huge, it is less substantial than the improvements in the large architecture.

We analyze the project-wise results for a detailed comparison. In the case of methods that have similar aggregated metrics, the comparisons vary between different projects. For example, \adapter{} outperforms \full{} in Kibana while the results are reversed in Qooxdoo. In contrast, for the methods with a huge difference in aggregated metrics, the conclusions also hold in project-wise analysis. For example, \full{} outperforms Default in all projects (Using TFix-Large). Therefore, while we cannot compare methods with similar results, we can make strong conclusions on the large differences. 

\full{} and \adapter{} are the best methods in project-wise analysis. We analyze 16 cases (8 for TFix-Small and 8 for TFix-Large). Using \full{}, the accuracy is improved in 13 cases, slightly reduced in 1 case, and remains the same in 2 cases. Using \adapter{}, the accuracy is improved in 11 cases, reduced in 2 cases, and remains the same in 3 cases. 5 of 8 cases that do not show improvement belong to ONM or LivelyKernel. ONM is an easy project. All of its bugs are fixed by TFix regardless of the design method or model size. LivelyKernel is the smallest project with the lowest number of adaptation and test samples. Therefore, DA methods' results are less reliable on LivelyKernel compared to the bigger projects. Finally, all DA methods show improvements over the Default in most cases. This shows the need for and benefit of using Domain Adaptation for TFix. 
 
Baseline has substantially lower effectiveness in TFix-Small compared to TFix-Large. The reason is that Baseline pays the same attention to all samples regardless of the domain they belong to. Consequently, a large model can automatically associate a part of its abundant parameters with a small domain and learn internal distributions. On the contrary, a small model does not have the capacity to recognize small domains inside the dataset and falls behind. Therefore, we conclude that full training with data of the target project can reach a high accuracy only if the model size is large enough, and it might not apply to a smaller model such as TFix-Small.
 
The results show that the best type of \curr{} is \currs{}. This type uses more software-relevant information because the curriculum is extracted from CodeBERT. Nonetheless, this method can not compete with the best methods in TFix-Large as the main architecture of TFix. The model size highly affects the improvements that vary from project to project. Therefore, we cannot say \curr{} is an effective, and generalizable method considering the three different types we implemented.

\full{} and \adapter{} showed the best effectiveness regardless of model size. They both improve the pretrained model on several target projects, including easy and hard ones. The improvement is substantial and holds for the two very different sizes of TFix. The difference between \full{} and \adapter{} is low and depends on the characteristics of the target project. Therefore, we say none of them is superior to another.

Finally, we conclude that the proposed DA framework is highly effective in improving accuracy. \full{} and \adapter{} are the best methods. They show the best improvement when applied to the large architecture size.

\paragraph{RQ2.2) \rqbx}
\begin{table}[]
\caption{Model Preparation Time for TFix-Small and TFix-Large}
\label{tab:model_preparation_time}
\centering
\begin{tabular}{lll}
  Method          & TFix-Small   & TFix-Large       \\
  \toprule
Default                                               & 0s         & 0s             \\
Baseline                                                & 6h 38m 48s & 1d 19h 19m 10s \\
\full {}                                           & 0m 38s     & 7m 09s         \\
\adapter{} & 2m 46s     & 5m 23s         \\
\currs{}           & 30s        & 1m 31s         \\
\currc{}         & 8s         & 0m 47s         \\
\currl{}             & 9s         & 1m 0s      \\
\bottomrule
\end{tabular}
\end{table}

In this section, we analyze the efficiency of each domain adaptation method. Table \ref{tab:model_preparation_time} shows time needed to prepare each model. We adapt the model in each of 8 target projects and report the average time for each domain adaptation method. 

Default has the most efficient training time as it requires no extra effort. We take the pretrained model as-is and use it for the target projects. Baseline has the worst preparation time because it trains the model on the general corpus. This preparation time is lengthy in TFix-Large as the model is bigger and takes more time at each feed-forward and backpropagation. This might make this approach entirely impractical according to the available resources.

The proposed DA methods all have a similar scale of preparation time. It takes around a minute to adapt TFix-Small on a target project and around a few minutes to adapt TFix-Large. \curr{} group has a slightly better efficiency than \full{}. They are trained similarly to \full{}, but the change in the training scheme leads to faster convergence with fewer epochs. The extra cost of re-ordering samples is negligible due to our efficient implementation. We calculate the length of samples or their distance to the general corpus only once in \currl{} and \currs{}. \currs{} takes relatively longer as the program needs to create vectors of samples to find distances. The advantage of using \curr{} compared to other methods is for the case when the models have to be trained very fast. For example, if each developer has a separate model on his machine and if frequent adaptation is required.

\adapter{} optimizes a small portion of parameters and freezes the rest of the network. However, the number of epochs needed to optimize the randomly initialized adapter layers might be more. This might take more optimization time in total. For example, \adapter{} takes 4.36 times more preparation time rather than \full{} for TFix-Small. On the other hand, the efficiency of \adapter{} is slightly better in TFix-Large because the size of the adapter layers is tiny compared to the whole model. Thus, even more epochs of their optimization take less time than a few epochs of optimizing the full network.

Overall, the differences between preparation times of DA methods are so low that does not affect the decision of choosing the best one in normal conditions. However, suppose one uses a couple of models for different domains, or separate models are used for several users, or data changes so fast and needs frequent retraining. In such cases, the differences become more important. For example, \curr{} methods with TFix-Small are better choices when models need to be prepared in less than 10 seconds.

\begin{table}[]
\caption{Model Sizes for TFix-Small and TFix-Large}
\label{tab:model_sizes}
\centering
\begin{tabular}{lcc}
Module             & TFix-Small   & TFix-Large   \\
\toprule
                            Base model & 232MB      & 2.8GB      \\
                            Each lightweight adapter layer                                       & 212K (*12) & 416K (*48) \\
                            \midrule
 Fine Tuning with Adapter Layer                                       & 235MB      & $\sim$2.8G
\end{tabular}
\end{table}

\begin{table}[]
\caption{Inference Times for TFix-Small and TFix-Large}
\label{tab:iference_time}
\centering
\begin{tabular}{ccc}
\multicolumn{1}{l}{}                & TFix-Small   & TFix-Large   \\
\midrule
Inference Time                                                                 & 0.39 sec   & 1.06 sec  
\end{tabular}
\end{table}

Model sizes are reported in Table \ref{tab:model_sizes}. This is the space that the model takes on the hard disk. This is highly correlated with the run-time storage it occupies, such as memory or GPU. These numbers are independent of the project on which the model is adapted and are based on the default saving function used in the original study of TFix.

The base model is the whole network used in Default, Baseline, \full{} and \curr{}. In \adapter{}, lightweight adapter layers are added to the base model. For example, 12 adapter layers are used in TFix-Small, and 48 adapter layers are used in TFix-Large based on the number of layers in the core transformers. Therefore, the size of \adapter{} is the sum of the base model size and the total size of all adapter layers added to it. Table \ref{tab:model_sizes} proves the extra usage of \adapter{} is very low, and therefore, we can conclude that all methods have almost the same run-time usage and disk storage requirements. Therefore, the required capacity is low, and the added volume in \adapter{} does not make this method impractical.

In the case of multi-domains, the difference is extremely more, though. Using the default approach or \adapter{}, one needs to keep one base model and one set of adapter layers for each domain. Otherwise, one needs to keep a full model for each domain of interest and load it to memory if models are used simultaneously. So, in this case, \adapter{} becomes incredibly more efficient.

Inference times are reported in Table \ref{tab:iference_time}. Inference time is the same between all methods. Using domain adaptation does not change it because the input passes the same path in the model each time. There is an extra cost of inference time in \adapter{}. However, it is so low that it does not change the average. The overall inference time is around a second, even in large architecture, making the method practical. That is, adapted models can fix bugs almost in real-time.

We conclude that the DA framework is practical in terms of efficiency according to the preparation time, model size, and inference time. The efficiency in the preparation time is a noteworthy advantage compared to Baseline. The domain adaptation methods can also be selected based on the end user's requirements, such as the need for real-time fixes or multi-domain DA.

\paragraph{RQ2.3) \rqby}

\begin{table}[]
\caption{Exposure Bias Exact Match Accuracy of Domain Adapted TFix-Small Models}
\label{tab:exposure_small}
\centering
\begin{tabular}{ccccccc}
Project      & \fullm{}  & \adapterm{} & \currsm{} & \currlm{} & \currcm{} \\
\midrule
Qooxdoo      & 27.00 & 29.02   & 34.00    & 35.02  & 35.02   \\
Kibana       & 34.00 & 35.02   & 33.01    & 35.02  & 36.02   \\
Ember.js     & 32.02 & 35.02   & 34.00    & 35.02  & 35.02   \\
Core-js      & 34.00 & 35.02   & 35.02    & 35.02  & 35.02   \\
ONM          & 34.00 & 33.01   & 35.02    & 35.02  & 35.02   \\
Sequelize    & 31.00 & 35.02   & 35.02    & 35.02  & 35.02   \\
Dcos-ui      & 34.00 & 35.02   & 35.02    & 35.02  & 35.02   \\
LivelyKernel & 34.00 & 34.00   & 35.02    & 35.02  & 35.02   \\
\midrule
\multicolumn{1}{c}{Average}         & 34.00 & 34.68   & 34.51    & 35.02  & 35.15   \\
\multicolumn{1}{c}{Median}          & 34.00 & 35.02   & 35.02    & 35.02  & 35.02   \\
\midrule
\multicolumn{1}{c}{Pretrained TFix-Small}         & \multicolumn{5}{c}{38.03}     \\
\bottomrule
\end{tabular}
\end{table}

\begin{table}[]
\caption{Exposure Bias Exact Match Accuracy of Domain Adapted TFix-Large Models}
\label{tab:exposure_large}
\centering
\begin{tabular}{ccccccc}
Project      & \fullm{}  & \adapterm{} & \currsm{} & \currlm{} & \currcm{} \\
\midrule
Qooxdoo      & 45.03 & 43.02   & 41.01    & 48.00  & 48.00   \\
Kibana       & 47.01 & 44.01   & 48.00    & 47.01  & 47.01   \\
Ember.js     & 46.02 & 44.01   & 44.01    & 48.00  & 48.00   \\
Core-js      & 47.01 & 45.03   & 47.01    & 48.00  & 47.01   \\
ONM          & 46.02 & 47.01   & 48.00    & 48.00  & 48.00   \\
Sequelize    & 46.02 & 44.01   & 47.01    & 48.00  & 47.01   \\
Dcos-ui      & 48.00 & 45.03   & 49.00    & 48.00  & 48.00   \\
LivelyKernel & 46.02 & 44.01   & 48.00    & 48.00  & 48.00   \\
\midrule
\multicolumn{1}{c}{Average}         & 46.39 & 44.52   & 46.51    & 47.88  & 47.63   \\
\multicolumn{1}{c}{Median}          & 46.02 & 44.01   & 47.51    & 48.00  & 48.00   \\
\midrule
\multicolumn{1}{c}{Pretrained TFix-Large}         & \multicolumn{5}{c}{40.01}  \\
\bottomrule
\end{tabular}
\end{table}

Table \ref{tab:exposure_small} and Table \ref{tab:exposure_large} show the accuracy of adapted models on the 5000 samples selected from the source set. Each row refers to a project on which the pretrained model is adapted. Each column refers to the used DA method. All models and methods are tested on the same dataset. Average and Median are used as the aggregated metrics. The accuracy of the pretrained model on the same dataset is also reported in the last row for comparisons.

According to Table \ref{tab:exposure_small}, there is a drop in the accuracy of TFix-Small after adaptation. However, the drop is not catastrophic. On the contrary, the accuracy of TFix-Large slightly increases in all DA methods according to Table \ref{tab:exposure_large}. The differences are insignificant between different rows and columns for TFix-Small and TFix-Large.

When TFix-Small is adapted to a specific domain, it slightly forgets general patterns according to the results. This is because the model's capacity is limited. It cannot remember the general information while learning more specific ones. On the contrary, the large model has abundant parameters. They enable Tfix to learn new patterns in the target project without losing their effectiveness on the general data. TFix performs even better on the general dataset after adaptation. This is due to the extra information the adaptation samples provide. These conclusions are independent of the domain in which the model is adapted and held for all projects.

Among all proposed methods, the \curr{} group is slightly more potent in terms of exposure bias. The improvement is more in TFix-Large, and the loss is less in TFix-Small. The reason is that they gradually change the model. This helps the model avoid sudden changes in the network. However, differences are low. This prevents us from concluding that one method is superior to others.

All in all, the DA framework and the DA methods are reliable regarding Exposure Bias. The model does not overfit the target project, and the results are reliable. One explanation is that the APR data of TFix does not differ significantly between different projects. The newly added data does not contrast with the general one. It only feeds extra information to the model without making it forget the old one. That is why a large model like TFix-Large improves after DA. We mark it as an essential characteristic of domain shift in TFix data discovered in this study.

\vspace{5mm}
\begin{mdframed}[style=mpdframe,frametitle=RQ2 Summary]
The proposed domain adaptation framework shows excellent improvements compared to the default and baseline approaches on both TFix-Small and TFix-Large. Overall, \full{} and \adapter{} are the best methods. They are more effective if applied on TFix-Large. We state that the results are practical in terms of efficiency, and the extra cost of the framework is affordable. We show the possibility of using DA in personalized APR models in real-time and multi-domain contexts. We show that the adapted models and results are reliable regarding exposure bias. The final choice between DA methods depends on the use case and its requirements, including the number of domains and resources.
\end{mdframed}

\subsection{RQ3 Evaluation) \rqc}

\subsubsection{Design}

\begin{figure}[h]
  \centering
  \includegraphics[width=\linewidth]{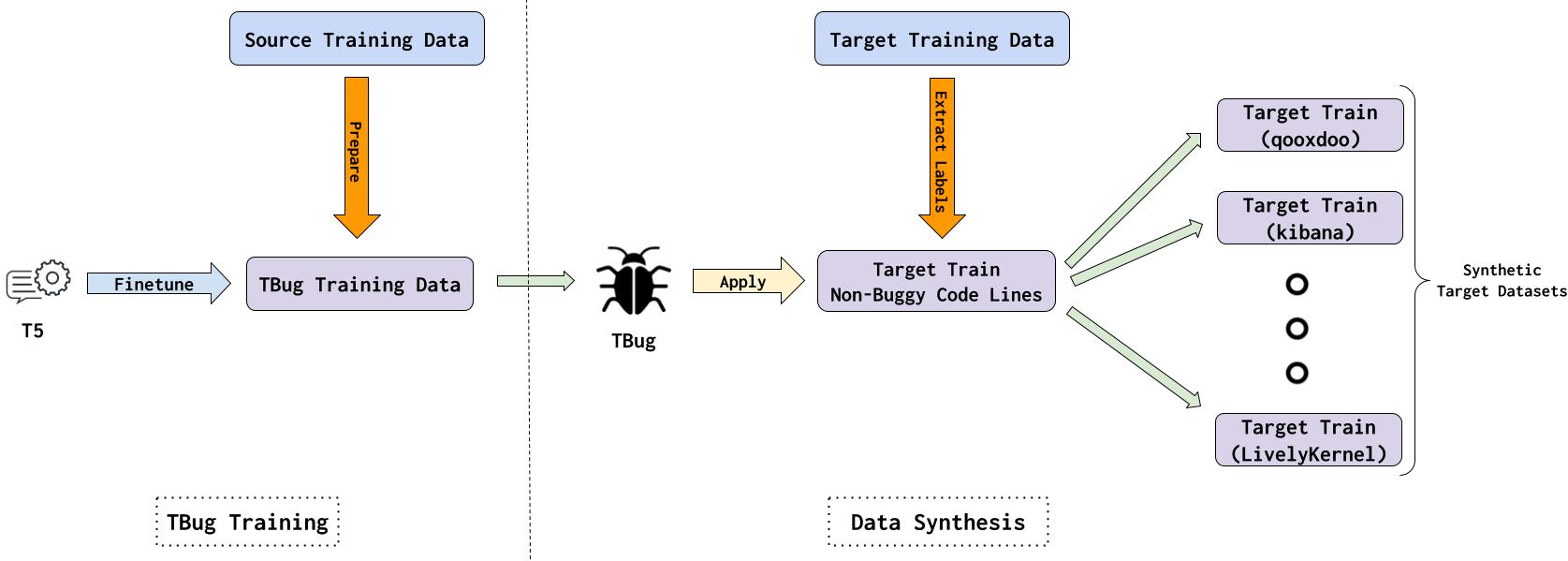}
  \caption{Experiment Design of TBug Training and Data Synthesis for Target Projects}
  \label{fig:rq3}
\end{figure}

In this research question, we make a semi-supervised assumption. That means only unlabeled data is provided in target projects. Therefore, we should not use any labeled data of the target set during adaptation. The unlabeled data is the code lines of target projects. We note that labeled data is still provided in the source set.


The experiment design is depicted in Figure \ref{fig:rq3}. We prepare the training data for TBug out of the source training dataset; We pick source-train and take its fixed lines as non-buggy lines. We add the code context (the line before and after the error line) to each sample and use the error type as metadata. This is the input to TBug. We also use buggy lines without context and metadata as the label. We call the resulting set ``TBug Training Data''. We train T5 on this data and the resulting model is TBug.

Next, we take target-train and extract labels out of it. These labels are non-buggy code lines. We remove their corresponding bugs consistent with the semi-supervised assumption of this experiment. We apply TBug on the non-buggy lines to generate bugs. Then, we create the bug-fix pairs and split them based on the projects. These splits are project-specific synthetic APR datasets.

In the last step, we replace the target set in RQ2 with the created datasets and repeat the DA framework. We choose \full{} as the DA method based on the results of RQ2. Finally, we test the adapted models on target-test that has never been used during previous steps of this RQ. Target-test is the same test data as all previous RQs. This makes the results comparable to the other methods and approaches. We compare the results with Default because it has the same semi-supervised assumption. Unlike other methods of RQ2, Default needs no access to the labeled data.

We repeat experiments with both T5-Small and T5-Large models to generalize the results. We call the trained models TBug-Small and TBug-Large, respectively. The choice between TBug-Small and TBug-Large to create a synthetic adaptation dataset is independent of the choice between TFix-Small and TFix-Large to fix the test bugs. For example, one might use TBug-Large to create synthetic data and TFix-Small to fix the test bugs. Nevertheless, we use the same architecture size for both because, typically, the decision of sizes depends on available resources such as time and memory. Therefore, the selected sizes are expected to be the same.

\subsubsection{Results}
\paragraph{RQ3.1) \rqcw}

\begin{table}[]
\caption{TBug Accuracy per Error Type}
\label{tab:TBug}
\centering
\footnotesize
\begin{tabular}{cccccc}
\multicolumn{1}{l}{}            & \multicolumn{2}{|c|}{Without MetaData} & \multicolumn{2}{c|}{With MetaData} &               \\
\midrule
Error Types                        & T5-small         & T5-large          & T5-small        & T5-large        & count all    \\
\toprule
\textbf{no-invalid-this}        & 1.24             & 0.62              & 2.48            & 5.59            & 161           \\
\textbf{no-throw-literal}       & 93.1             & 78.74             & 90.8            & 92.53           & 174           \\
\textbf{no-new-wrappers}        & 0                & 0                 & 33.33           & 66.67           & 3             \\
\textbf{guard-for-in}           & 10               & 5                 & 15              & 15              & 20            \\
\textbf{no-new-object}          & -                & -                 & -               & -               & 0             \\
\textbf{comma-style}            & 6.49             & 10.71             & 22.4            & 23.7            & 308           \\
\textbf{prefer-spread}          & 66.28            & 74.42             & 66.28           & 77.91           & 86            \\
\textbf{no-caller}              & 0                & 0                 & 5.26            & 21.05           & 19            \\
\textbf{no-extra-bind}          & 0                & 0                 & 0               & 0               & 2             \\
\textbf{no-array-constructor}   & 0                & 0                 & 0               & 0               & 1             \\
\textbf{prefer-rest-params}     & 61.29            & 5.16              & 61.94           & 63.87           & 155           \\
\textbf{generator-star-spacing} & -                & -                 & -               & -               & 0             \\
\textbf{no-this-before-super}   & -                & -                 & -               & -               & 0             \\
\textbf{no-extend-native}       & 0                & 0                 & 41.67           & 54.17           & 24            \\
\textbf{no-undef}               & 0                & 0                 & 3.1             & 4.65            & 129           \\
\textbf{no-useless-escape}      & 0                & 0                 & 0               & 9.09            & 11            \\
\textbf{no-dupe-keys}           & 0                & 2.94              & 26.47           & 32.35           & 34            \\
\textbf{no-console}             & 0                & 0                 & 16.67           & 33.33           & 12            \\
\textbf{no-constant-condition}  & 0                & 0                 & 0               & 25              & 12            \\
\textbf{no-duplicate-case}      & 0                & 50                & 50              & 50              & 2             \\
\textbf{no-empty}               & 9.09             & 9.09              & 54.55           & 45.45           & 11            \\
\textbf{no-extra-semi}          & 69.67            & 71.72             & 74.59           & 77.87           & 244           \\
\textbf{no-redeclare}           & 24.44            & 17.78             & 28.89           & 40              & 45            \\
\textbf{no-cond-assign}         & 30               & 10                & 30              & 20              & 10            \\
\textbf{no-extra-boolean-cast}  & 35               & 25                & 55              & 55              & 20            \\
\textbf{no-fallthrough}         & -                & -                 & -               & -               & 0             \\
\textbf{no-unreachable}         & 22.22            & 27.78             & 27.78           & 27.78           & 18            \\
\textbf{valid-typeof}           & 0                & 0                 & 0               & 0               & 1             \\
\textbf{no-unsafe-finally}      & 0                & 0                 & 0               & 0               & 1             \\
\textbf{no-unused-vars}         & 0                & 0                 & 0               & 1.61            & 62            \\
\textbf{no-debugger}            & 24.79            & 17.95             & 55.56           & 54.7            & 117           \\
\textbf{no-unsafe-negation}     & 0                & 0                 & 0               & 0               & 2             \\
\textbf{no-case-declarations}   & 100              & 100               & 100             & 100             & 1             \\
\textbf{no-self-assign}         & 0                & 0                 & 100             & 100             & 2             \\
\textbf{no-process-exit}        & 100              & 50                & 100             & 100             & 2             \\
\textbf{no-inner-declarations}  & 0                & 0                 & 50              & 50              & 4             \\
\textbf{for-direction}          & 0                & 0                 & 0               & 0               & 1             \\
\textbf{no-compare-neg-zero}    & -                & -                 & -               & -               & 0             \\
\textbf{no-sparse-arrays}       & -                & -                 & -               & -               & 0             \\
\textbf{no-func-assign}         & 0                & 0                 & 50              & 50              & 2             \\
\textbf{no-const-assign}        & -                & -                 & -               & -               & 0             \\
\textbf{no-global-assign}       & 0                & 0                 & 0               & 0               & 8             \\
\textbf{use-isnan}              & 0                & 0                 & 0               & 100             & 1             \\
\textbf{no-unused-labels}       & 0                & 0                 & 0               & 0               & 1             \\
\textbf{require-yield}          & -                & -                 & -               & -               & 0             \\
\textbf{getter-return}          & 0                & 0                 & 0               & 100             & 1             \\
\textbf{no-dupe-class-members}  & -                & -                 & 0               & -               & 0             \\
\textbf{no-ex-assign}           & -                & -                 & -               & -               & 0             \\
\textbf{constructor-super}      & 0                & 0                 & 0               & 0               & 1             \\
\textbf{no-new-symbol}          & 0                & 0                 & 0               & 0               & 1             \\
\textbf{no-empty-pattern}       & 0                & 0                 & 0               & 100             & 1             \\
\textbf{no-class-assign}        & -                & -                 & -               & -               & 0   \\
\midrule
\textbf{Weighted Average}       & \textbf{33.1}    & \textbf{27.13}    & \textbf{41.4}   & \textbf{44.68}  & \textbf{1710}
\end{tabular}
\end{table}

Table \ref{tab:TBug} reports the exact match accuracy of TBug. This is measured on target-train that includes 1710 code samples; therefore, some rare bug types have no representative instances. We expect these bug types to be rare in the test dataset of DA methods because the same algorithm extracts them.

Results are reported for two types; ``Without MetaData'' and ``With MetaData''. The former refers to the setting where the error types are not leveraged, and the latter refers to the main version of the proposed TBug method.

The effectiveness of TBug significantly increases when the bug types are provided to the model. The accuracy of TBug-Small increases from 33.1\% to 41.4\%, and the accuracy of TBug-Large increases from 27.13\% to 44.68\%. This is because, without access to the bug types, the model generates bugs that are accurate but different from the references. This confuses the model and makes it update the network in the wrong order during backpropagation. However, error type information helps the model understand how to generate bugs based on bug types differently. This improvement can have a substantial side effect on the downstream tasks that use synthetic datasets, including domain adaptation for TFix. Therefore, we include bug type in the implementation for RQ3.2.

The effectiveness of TBug varies between different error types. For example, it is hard for TBug to create ``no-invalid-this'' while it is relatively easy to create ``no-throw-literal''. A similar fact holds for TFix as reported in \cite{berabi2021tfix}, but the numbers are different. Some error types that are easy to fix by TFix are hard to generate by TBug and vice versa. 

The accuracy of TBug-large is 44.68\%. That means 44.68\% of generated bugs are identical to the referenced bugs in the dataset. The remaining 55.32\% of the generated bugs might still be helpful. They might be buggy code lines that are different from references. We use all of the generated bugs for RQ3.2 and leave filtering incorrect bugs as future work.

Finally, the accuracy of TBug-Large is slightly more than that of TBug-Small, with a difference of 3.28\%. However, the difference is not huge and varies based on the error type. We say that the choice of architecture size in TBug depends on the available computation resources. Therefore, we use TBug-Small to create bugs for TFix-Small and TBug-Large to create bugs for TFix-Large in RQ3.2.



\paragraph{RQ3.2) \rqcx}
We report the final results of DA with synthetic data of TBug in Table \ref{tab:proposed_small} and Table \ref{tab:proposed_large}. Former, reports the accuracy of TFix-Small adapted on dataset generated by TBug-Small. Latter, reports the accuracy of TFix-Large adapted on the dataset generated by TBug-Large. The highest number is marked in each row.

Using the small T5 size, the effectiveness improves in 4 projects, remains the same in two projects, and reduces in two projects. On average, the accuracy improves 3.96\%. Using the large T5 size, accuracy improves in 5 projects, remains the same in 3 projects, including the easy ONM, and increases 5.66\% on average.

The results show that synthetic data generated by TBug are of high quality for our downstream task. They improve the effectiveness of TFix in case no labeled data exists for target projects. The improvements are notable regardless of the model size. This extends our DA framework to semi-supervised conditions. That means the DA framework still adapts the models on project-specific data if no labeled data is provided for target projects. This is a significant achievement to remove a supervised assumption from the proposed framework.

\begin{table}[]
\caption{Comparison between the Default Approach and \full{} with Synthetic Data of TBug using Small Architectures}
\label{tab:proposed_small}
\centering
\begin{tabular}{lccc}
Project      & \multicolumn{1}{l}{\# Samples} & \multicolumn{1}{l}{Default} & \multicolumn{1}{l}{\fullm{} with Synthetic Data} \\
\toprule
Qooxdoo      & 547                            & 48.70                       & \textbf{51.30}                                 \\
Kibana       & 231                            & \textbf{43.14}                       & 41.18                                 \\
Ember.js     & 218                            & 51.06                       & \textbf{57.45}                                 \\
Core-js      & 201                            & 21.95                       & \textbf{51.22}                                 \\
ONM          & 194                            & 100                         & 100                                   \\
Sequelize    & 180                            & 40.00                       & \textbf{40.00}                                 \\
Dcos-ui      & 166                            & \textbf{94.60}                       & 86.49                                 \\
LivelyKernel & 159                            & 61.11                       & \textbf{66.67}                                 \\
\midrule
\multicolumn{2}{l}{Weighted Average}          & 54.93                       & \textbf{58.87}                                 \\
\multicolumn{2}{l}{Average}                   & 57.57                       & \textbf{61.79}                                 \\
\multicolumn{2}{l}{Median}                    & 49.88                       & \textbf{54.38}   \\                             
\bottomrule
\end{tabular}
\end{table}

\begin{table}[]
\caption{Comparison between the Default Approach and \full{} with Synthetic Data of TBug using Large Architectures}
\label{tab:proposed_large}
\centering
\begin{tabular}{lccc}
Project      & \multicolumn{1}{l}{Samples} & \multicolumn{1}{l}{Default} & \multicolumn{1}{l}{\fullm{} with Synthetic Data} \\
\toprule
Qooxdoo      & 547                            & 46.08                       & \textbf{48.70}                                 \\
Kibana       & 231                            & 45.10                       & \textbf{47.06}                                 \\
Ember.js     & 218                            & 51.06                       & \textbf{63.83}                                 \\
Core-js      & 201                            & 39.02                       & \textbf{46.34}                                 \\
ONM          & 194                            & 40.00                       & 40.00                                 \\
Sequelize    & 180                            & 100                         & 100                                   \\
Dcos-ui      & 166                            & 67.57                       & \textbf{94.59}                                 \\
LivelyKernel & 159                            & 66.67                       & 66.67                                 \\
\midrule
\multicolumn{2}{l}{Weighted Average}          & 54.19                       & \textbf{59.85}                                 \\
\multicolumn{2}{l}{Average}                   & 56.94                       & \textbf{63.40}                                 \\
\multicolumn{2}{l}{Median}                    & 48.58                       & \textbf{56.27}               \\
\bottomrule
\end{tabular}
\end{table}

\vspace{5mm}
\begin{mdframed}[style=mpdframe,frametitle=RQ3 Summary]
The proposed TBug method is effective in APR data synthesization. It successfully incorporates into the supervised domain adaptation framework of TFix. It removes the supervised assumption and improves its effectiveness on the projects that lack labeled data.
\end{mdframed}

`

\subsection{RQ4 Evaluation) \rqd}

\subsubsection{Design}
We pick the projects with at least 50 samples in the Java dataset of CodeXGLUE. We argue that the experiments and methods cannot be properly analyzed on projects with fewer samples for testing. This is the same argument as the target project selection in RQ1 with more relaxed constraints. The resulting set is 6 projects from the small dataset and 5 from the medium dataset. We use them all, label them as ``Target Set'' and leave the remaining as ``Source Set''. For each project, we split it into train, validation, and test

We relax constraints (use less intense filters) because the dataset of CodeXGLUE is smaller and more limited. If we use the same filters as RQ1, there would not be enough projects for experiments. We understand that the projects that have only a few samples might show relatively poor results.

We repeat the key points of RQ[1-3]. In RQ4.1, we train the model in two similar scenarios named Included and Excluded as RQ1. We measure the sensitivity of CodeXGLUE to do damage of domain. In RQ4.2, we domain adapt the pretrained CodeXGLUE model on each target project. We use the \full{} method for this RQ as no other method showed superiority to it in RQ2. 

In RQ4.3, we experiment with the effectiveness of CodeXBUG. We take the Encoder-Decoder architecture, put CodeBERT as the encoder, and randomly initialize the decoder according to CodeXGLUE training. Then, we pick the Java dataset of CodeXGLUE and swap the buggy and fixed lines. We train CodeXBUG on this dataset. Finally, we use CodeXBUG to generate the APR dataset for each project. We domain adapt CodeXGLUE on generated synthetic dataset on target projects.

We repeat all processes on both small and medium datasets of CodeXGLUE to add to the generalizability of our findings.

\subsubsection{Results}
\paragraph{RQ4.1) \rqdw}
Table \ref{tab:codexglue_comparison_included_excluded_small} and Table \ref{tab:codexglue_comparison_included_excluded_medium} show the accuracy of CodeXGLUE in Included and Excluded designs on small and medium datasets, respectively. Each cell is the exact match accuracy between generated fixes and the references. Weighted Average, Average, and Median are used as the aggregated metrics. Higher numbers are marked in each row.

In the included design, 1 project in the small set and 3 of 5 projects in the medium set have 0\% accuracy. It means CodeXGLUE cannot fix any of the errors in these 4 cases. The reason is that CodeXGLUE is a weaker method compared to TFix, especially when applied to samples with longer lengths (Medium Dataset). 

Other than these four cases, the accuracy of projects is dropped on 3 of 4 projects in the small set and 1 of 2 projects in the medium set when source projects are excluded. The accuracy remains unchanged in the other two projects. The weighted average drops from 31.53\% to 05.40\% (Small Dataset) and from 06.38\% to 01.06\% (Medium dataset). These drops include the results of projects that have 0\% accuracy. Snobot2015 and Orientdb are the projects that suffer severely from domain shift. CodeXGLUE cannot fix any samples of these two projects if it has not seen other examples in the training dataset. These all confirm that CodeXGLUE is sensitive to Domain Shift. However, this threat is less visible compared to TFix-Large.

\begin{table}[]
\caption{Exact Match Accuracy of the Included and Excluded Experiments on Target Projects of the Small Dataset with CodeXGLUE}
\label{tab:codexglue_comparison_included_excluded_small}
\centering
\begin{tabular}{cccc}
Project              & \# Samples    & Included & Excluded \\
\midrule
Snobot 2015          & 170           & \textbf{85.29}    & 0.00     \\
OrientDB             & 105           & 4.76     & 4.76     \\
Coprhd Controller    & 104           & \textbf{9.52}     & 4.76     \\
Chromium             & 58            & \textbf{16.67}    & 8.33     \\
IHMC                 & 56            & 0        & \textbf{8.33}     \\
RStudio              & 54            & 9.09     & \textbf{18.18}    \\
\midrule
\multicolumn{2}{c}{Weighted Average} & \textbf{31.53}    & 05.40    \\
\multicolumn{2}{c}{Average}          & \textbf{20.80}    & 07.39    \\
\multicolumn{2}{c}{Median}           & \textbf{09.30}    & 06.54    \\
\bottomrule
\end{tabular}
\end{table}

\begin{table}[]
\caption{Exact Match Accuracy of the Included and Excluded Experiments on Target Projects of the Medium Dataset with CodeXGLUE}
\label{tab:codexglue_comparison_included_excluded_medium}
\centering
\begin{tabular}{lccc}
Project           & \multicolumn{1}{l}{\# Samples} & \multicolumn{1}{l}{Included} & \multicolumn{1}{l}{Excluded} \\
\midrule
Coprhd Controller & 170                            & 0                            & 0                            \\
OrientDB          & 110                            & \textbf{22.73}                        & 0                            \\
Wordpress android & 62                             & 7.69                         & 7.69                         \\
IHMC              & 61                             & 0                            & 0                            \\
UIcase            & 58                             & 0                            & 0                            \\
\midrule
\multicolumn{2}{c}{Weighted Average}               & \textbf{06.38}                        & 01.06                        \\
\multicolumn{2}{c}{Average}                        & \textbf{06.08}                        & 01.54                        \\
\multicolumn{2}{c}{Median}                         & 0                            & 0                           \\
\bottomrule
\end{tabular}
\end{table}

\paragraph{RQ4.2) \rqdx}
In Table \ref{tab:codexglue_da_small} and Table \ref{tab:codexglue_comparison_included_excluded_medium}, we report the accuracy of three possible approaches. Default, Baseline, and Domain Adaptation. Like RQ2, the default approach (Default) is equivalent to the excluded design, and the baseline approach (Baseline) is equivalent to the Included design. Therefore, we borrow the results from RQ1. In each row, the highest number is marked.

In a small dataset, Default has the least effectiveness, with an average of 05.40\%. Domain Adaptation is the highest effective method, with an accuracy of 45.05\%. Baseline places in between with an accuracy of 31.53\%. The advantages of Domain Adaptation hold in the project-wise analysis as well. In all 6 projects, Domain Adaptation is more effective by a significant margin compared to Default and Baseline. However, Baseline does not hold superiority over Default in project-wise analysis. While the effectiveness jumps higher in Snobot 2015, it slightly improves in one project, remains the same in two projects, and reduces in two projects. This shows Baseline is not as effective and generalizable as Domain Adaptation.

In the medium dataset, numbers are lower because the longer samples are harder to fix for CodeXGLUE. It makes the results less reliable. For example, the accuracy of Coprhd Controller is 0\%  in all three methods. That means CodeXGLUE cannot fix any sample regardless of the approach. Domain Adaptation hits the highest accuracy in 3 of other 4 projects. On average, it reaches 15.96\% which is far above the accuracy of both Default and Baseline. Baseline improves the accuracy in one project, and does not change in the other 4 projects. This shows Baseline is not generalizable to the medium dataset.

These results show that Domain Adaptation with \full{} is extendible to CodeXGLUE. It improves the accuracy in 10 of 11 cases without modifying the framework. Results also show Baseline is not effective compared to Domain Adaptation for CodeXGLUE.

\begin{table}[]
\caption{Comparison Between Default and FullFineTuning with Synthetic Data of CodeXBUG, on the Small Dataset}
\label{tab:codexglue_da_small}
\centering
\begin{tabular}{lcccc}
Project              & \# Samples    & Default                   & Baseline                  & Domain Adaptation (\fullm{})     \\
\midrule
Snobot 2015          & 170           & 0.00                      & 85.29                     & \textbf{97.06}                     \\
OrientDB             & 105           & 4.76                      & 4.76                      & \textbf{78.64}                     \\
Coprhd Controller    & 104           & 4.76                      & 9.52                      & \textbf{78.15}                     \\
Chromium             & 58            & 8.33                      & 16.67                     & \textbf{72.06}                     \\
IHMC                 & 56            & 8.33                      & 0                         & \textbf{60.65}                     \\
RStudio              & 54            & 18.18                     & 9.09                      & \textbf{57.66}                     \\
\midrule
\multicolumn{2}{c}{Weighted Average} & \multicolumn{1}{c}{05.40} & \multicolumn{1}{c}{31.53} & \multicolumn{1}{c}{\textbf{45.04}} \\
\multicolumn{2}{c}{Average}          & \multicolumn{1}{c}{07.39} & \multicolumn{1}{c}{20.80} & \multicolumn{1}{c}{\textbf{33.50}} \\
\multicolumn{2}{c}{Median}           & \multicolumn{1}{c}{06.54} & \multicolumn{1}{c}{09.30} & \multicolumn{1}{c}{\textbf{17.43}}\\
\bottomrule
\end{tabular}
\end{table}

\begin{table}[]
\caption{Comparison between Default and FullFineTuning with Synthetic Data of CodeXBUG, on the Medium Dataset}
\label{ref:tab:codexglue_da_medium}
\centering
\begin{tabular}{lcccc}
Project              & \# Samples    & Default                   & Baseline                  & Domain Adaptation (\fullm{})                     \\
\midrule
Coprhd Controller    & 170           & 0                         & 0                         & 0                         \\
OrientDB             & 110           & 0                         & 22.73                     & \textbf{40.91}                     \\
Wordpress android    & 62            & \textbf{7.69}                      & \textbf{7.69}                      & 0                         \\
IHMC                 & 61            & 0                         & 0                         & \textbf{7.69}                      \\
UIcase               & 58            & 0                         & 0                         & \textbf{41.67}                     \\
\midrule
\multicolumn{2}{c}{Weighted Average} & \multicolumn{1}{c}{01.06} & \multicolumn{1}{c}{06.38} & \multicolumn{1}{c}{\textbf{15.96}} \\
\multicolumn{2}{c}{Average}          & \multicolumn{1}{c}{01.54} & \multicolumn{1}{c}{06.08} & \multicolumn{1}{c}{\textbf{18.05}} \\
\multicolumn{2}{c}{Median}           & \multicolumn{1}{c}{0}     & \multicolumn{1}{c}{0}     & \multicolumn{1}{c}{\textbf{7.69}} \\
\bottomrule
\end{tabular}
\end{table}

\paragraph{RQ4.3) \rqdy}
Table \ref{tab:codexglue_aug_small} and Table \ref{tab:codexglue_aug_medium} show the results of using DA for CodeXGLUE. In this RQ, we use the synthetic data created by CodeXBUG. We compare this result with Default, which is the alternative approach for lack of data.

\begin{table}[]
\caption{Comparison between Default and FullFineTuning with Synthetic Data of CodeXBUG, on the Small Dataset}
\label{tab:codexglue_aug_small}
\centering
\begin{tabular}{lcccc}
Project           & \# Samples & \# Tests & Default                   & Data Synthetic \\
\midrule
Snobot 2015       & 170        & 34           & 0.00                      & 70.59             \\
OrientDB          & 105        & 21           & 4.76                      & 9.52              \\
Coprhd Controller & 104        & 21           & 4.76                      & 4.76              \\
Chromium          & 58         & 12           & 8.33                      & 8.33              \\
IHMC              & 56         & 12           & 8.33                      & 16.67             \\
RStudio           & 54         & 11           & 18.18                     & 18.18             \\
\midrule
\multicolumn{3}{c}{Weighted Average}          & \multicolumn{1}{c}{05.40} & 28.82             \\
\multicolumn{3}{c}{Average}                   & \multicolumn{1}{c}{07.39} & 21.34             \\
\multicolumn{3}{c}{Median}                    & \multicolumn{1}{c}{06.54} & 13.10         \\
\bottomrule
\end{tabular}
\end{table}

\begin{table}[]
\caption{Comparison between the Default Approach and FullFineTuning with Synthetic Data of CodeXBUG on the Medium Dataset}
\label{tab:codexglue_aug_medium}
\centering
\begin{tabular}{lcccc}
Project           & \# Samples & \# Tests & Default                   & Data Synthetic \\
\midrule
Coprhd Controller & 170        & 34           & 0                         & 0                 \\
OrientDB          & 110        & 22           & 0                         & 9.09              \\
Wordpress android & 62         & 13           & 7.69                      & 0                 \\
IHMC              & 61         & 13           & 0                         & 0                 \\
UIcase            & 58         & 12           & 0                         & 25.00             \\
\midrule
\multicolumn{3}{c}{Weighted Average}          & \multicolumn{1}{c}{01.06} & 05.32             \\
\multicolumn{3}{c}{Average}                   & \multicolumn{1}{c}{01.54} & 06.82             \\
\multicolumn{3}{c}{Median}                    & \multicolumn{1}{c}{0}     & 00.00       \\
\bottomrule
\end{tabular}
\end{table}

In the small dataset, the accuracy improves in 3 projects and remains the same in 3 others. The improvement is enormous in OrientDB and marginal in two others. In the medium dataset, we exclude Coprhd Controller as it showed 0\% accuracy in all experiments of RQ4.2. Other than that, the accuracy improves in two projects, remains the same in two projects, and reduces in one project. All in all, CodeXBUG shows improvements in 5 of 11 cases supporting the generalizability of our proposed method. However, this improvement is less significant than TBug. The main reason is that CodeXGLUE is a weaker tool compared to TFix. For some of the projects, it cannot fix most bugs with real data. Therefore, it cannot show the benefits of synthetic data in hard-to-fix projects. The other reason is that the dataset of CodeXGLUE is smaller than TFix's, which does not reveal the proposed methods' benefits. They have less target domain data to create synthetic bugs. \\

\begin{mdframed}[style=mpdframe,frametitle=RQ4 Summary]
Our experiments with CodeXGLUE support the generalizability of our key findings. Similar to TFix, CodeXGLUE is sensitive to domain shift issues. Its accuracy drops when target projects are excluded from the training data. Domain Adaptation with \full{} helps it recover the loss. It reaches higher effectiveness on each target project with a considerable margin. The idea of bug generation increases CodeXGLUE's effectiveness in case of no data. However, the improvement is less substantial due to the limitation of the dataset and the tool.
\end{mdframed}

\subsection{Additional Discussion}
This section discusses the points that did not fit inside the research questions' inter-discussions. 

First, some findings hold for all models or projects. For example, both TFix and CodeXGLUE benefit from domain adaptation using \full{} with the real data. This shows the effectiveness of the proposed framework using at least one DA method, which is the primary purpose of this study. On the other hand, some approaches or methods are limited to specific datasets or projects. For example, the baseline approach improves the accuracy in TFix-Large while it shows relatively less effectiveness in TFix-Small and CodeXGLUE. Therefore, we do not consider them a generalizable domain adaptation approach. 

The damage of domain shift to APR models is one of the things that is not the same between all methods and projects. In RQ1, we show that TFix-Large and TFix-Small react differently to domain shift. TFix-Large loses effectiveness while TFix-Small has relative resistance. Moreover, in RQ4.1, we see that the accuracy of CodeXGLUE reduces in some projects while it remains the same in others. We conclude that these APR models are sensitive to domain shift as their accuracy might drop. However, the actual drop depends on the size of the model and also the distribution of projects.

The effectiveness of the bug synthesis method is one of the other things that does not firmly hold on to all experiments. While the improvements are significant for TFix, they are specific to some projects on CodeXGLUE. The main reason is that the bug synthesis method needs a minimum amount of data, like all learning-based methods. The threshold is higher than the DA framework as the synthetic data might be of less quality than the actual data. This amount of data is not available for all projects of CodeXGLUE. Therefore, the conclusions are less intense. Overall, we consider the improvements in TFix and some projects of CodeXGLUE promising enough and encourage future work to continue in this area. For example, while we did not have access to more unlabeled data of projects according to the limitations of datasets, future studies might leverage each project's abundance of code lines in the real world to hit higher results. 

Second, \citeauthor{berabi2021tfix} \cite{berabi2021tfix} concluded that T5-large is the best architecture choice of TFix. However, they did not consider the effects of domain shift, which is the main motivation of this study. This is an important flaw because Domain Shift is not a rare or uncommon condition. In fact, TFix is expected to be tested on projects that are not part of the training set. This motivates us to review and refine the conclusions made by \citeauthor{berabi2021tfix} assuming the potential damage of domain shift.

The approach suggested by \citeauthor{berabi2021tfix} is to use TFix pretrained models on target projects without modifications, which is equal to this study's default approach. Table \ref{tab:comparison_excluded} shows that TFix-Large and TFix-Small both have similar effectiveness in the case of domain shift if the default approach is used. Therefore, we argue that in the real-world scenario and considering the approach of TFix's original paper, the large architecture of TFix has no advantage over small architecture in terms of effectiveness, which is contrary to the conclusions of \citeauthor{berabi2021tfix}. Besides, TFix-Small is a lighter model according to Table \ref{tab:model_sizes}, and Table \ref{tab:iference_time}. It takes 1/10 of the volume for both runtime storage, such as GPU or memory, and disk storage. Moreover, its inference time per instance is 63\% lower than TFix-Large, which is highly important in real-time systems. Therefore, we conclude that TFix-Small is even a better architecture choice if domain adaptation methods are not employed. While it has the same effectiveness, it is extraordinarily lighter, more efficient, and easier to maintain. This contradicts the conclusions of \citeauthor{berabi2021tfix}. This also further states the importance of considering the dangers of domain shift to software engineering models. 

However, suppose one is using our proposed domain adaptation framework. In that case, TFix-Large becomes a better choice in terms of effectiveness by a margin of 6.65\% in the weighted average according to Table \ref{tab:comparison_methods_large} and Table \ref{tab:comparison_methods_small}. This is one of the best achievements of this study to keep the advantage of the higher capacity of TFix-Large, which the original study could not do based on real-world assumptions. 

Still, the final decision about the model size is a trade-off between effectiveness and efficiency concerns. Considering all results of this work, one might decide to continue with the default approach, especially if the labeled data is not provided and efficiency is more important than the differences in effectiveness.

\subsection{Limitations of the Proposed Methods and the Conducted Study}
\label{general_limits}
The most important limitation of our study is the APR methods under experiment. One may argue that the results and findings might not be applicable to other APR methods, such as the ones introduced after our study. To address this problem, we replicated the study on two different sizes of TFix; TFix-Small and TFix-Large. The latter is the most potent architecture of TFix, and the former represents a smaller and more efficient model with lower capacity. In addition, we investigated the generalization of key findings in RQ4. We tested the system on CodeXGLUE, a different technique in model architecture, data processing, and other ideas. The experiments of CodeXGLUE were on a Java dataset with fundamentally different data points than the ones of TFix. This was our best to add to the generalizability of the findings.

In the experiments of CodeXGLUE, we have limitations for dataset quality. The average number of samples in the selected target projects is low. This issue is more intense in the dataset of CodeXGLUE. CodeXGLUE's dataset has fewer samples per project compared to TFix. For example, only one project has more than 150 samples in each dataset of CodeXGLUE, while this number is 24 for TFix. This makes the results less strong. The effect of domain shift and adaptation is less noticeable when only a few samples exist in the target domain. Thus, it is not surprising that the accuracy of ``Coprhd Controller'' is 0 in all approaches and methods. We note this as a limitation of this study. APR datasets have sufficient data, but their samples are split between many projects. Therefore, they lack projects with a high number of samples for each. This makes the results of the second method less intense. 

We addressed the lack of data issue by introducing the data synthesis method. The data synthesis generates APR samples for each project. In the real world, each project might come with an abundance of code lines that can be used to create large project-specific APR datasets. However, in our experiments, we did not have access to the pure source code of each project in addition to the APR dataset. The full datasets are not published. We mention it as a future direction to expand the experiments of the bug synthesis method and use them to create larger project-specific APR datasets. This might lead to higher effectiveness of the DA framework regardless of the available labeled data.

Another limitation is the accuracy metric. We used ``exact match'' as it is the metric used in all APR methods and is in common between them. We did not use the BLEU score or Code BLEU score. First, these metrics do not explicitly correlate with the software code's quality. Second, these metrics are shown to be correlated with the exact match score in related works. Likewise, \citeauthor{berabi2021tfix} did not use the BLEU score in their study. \citeauthor{berabi2021tfix} used another metric called ``Error Removal".  Error removal refers to the number of errors that are removed after modifications. We did not use these metrics for a couple of reasons. First and most important, the source code of this part of their study is not published in open source. Error removal is not a simple and straightforward metric. It is a complicated algorithm that cannot be replicated easily. For example, when a part of the code is changed, it might remove some errors and introduce other errors in the same location or other parts of the code. So, considering this change as an improvement is not trivial. The full algorithm is neither discussed in the original paper nor published elsewhere. Second, measuring the error removal score requires access to the full source code of projects as a requirement of ESLint. We do not have access to the full source codes as the authors did not publish them. Also, some of the projects are no longer freely available, making us unable to retrieve the source codes. Third, \citeauthor{berabi2021tfix} are the only ones who used Error Removal as an accuracy metric in all APR methods. Their results show that Error Removal is highly correlated with Exact Match. Therefore, we do not see a sheer necessity in adding this metric.

All in all, we consider the metrics used in this study as a limitation. However, this limitation is not only for our study but for APR research. We encourage future software research studies to work on software-specific metrics that measure the quality of generated code.

\subsection{Validity Threats}
In the following, we explain threats to validity in four sections, construct validity, internal validity, external validity, and conclusion validity.

\paragraph{Construct Validity} 
We should analyze the soundness of the metric. We check if the higher value of the metric means the higher quality of the APR model. Exact match measures if our prediction exactly matches the change made by the developer. This aligns with the APR method's purpose: to predict changes that developers will make.

Another threat to construct validity is that we assume the increase in the accuracy of the model in target projects is correlated with its capability to predict bug-fixes. However, the model might be simply overfitted on the small proportion of target data. We deny this threat in RQ2.3 when we measure the model's effectiveness on the original dataset and show it is not dropped.

\paragraph{Internal Validity} 
We use the same test set between different experiments of RQ1, RQ2.1, and RQ3 to ensure the changes are not due to the difference in the test dataset. 

One crucial threat to internal validity is that the drop in the accuracy of the excluded model in RQ1 is the result of a lesser amount of data in its training set. We argue that the portion of data excluded from the dataset in RQ1 is less than 2\% of the whole data. Therefore, such a huge drop cannot result from removing such a small portion of data.

Moreover, one can argue that the accuracy improvements in RQ2.1 might be due to the more data the model fed to the model. In other words, the model generally improves by getting new data, not by learning the target project. We argue that the amount of data that models are tuned on is little compared to the general dataset. Also, results of RQ2.3 show that the effectiveness of models is not increased hugely in the general dataset. They are reduced for TFix-Small and slightly improved for TFix-Large. The improvement of TFix large is marginal compared to the improvements on each target project. These all show that the effectiveness improvements are not only due to the increase in the number of samples. The model learns the distribution of each target project.

\paragraph{External Validity}
To address External Validity, first, we tested two sizes of TFix; TFix-Small and TFix-Large. Although they use the same method and dataset, they are hugely different regarding the number of learnable parameters. We also designed RQ4 to expand our findings on another APR method, CodeXGLUE. CodeXGLUE is a fundamentally different APR technique in terms of dataset, architecture, and method, which supports the generalization. 

\paragraph{Conclusion Validity}
We tested methods on 8 randomly selected projects for TFix and 11 randomly selected projects for CodeXGLUE. The selected projects are diverse and different than each other based on the number of samples and the accuracy of APR models on them. In the end, we explain in all research questions which conclusions hold in all projects, individually, and are aligned with changes in the aggregated metrics: weighted average, average, and median.

\section{Related Work}
\label{sc:related_work}
Domain adaptation has been cultivated in software engineering research. A few studies investigated DA to transfer models from high-resource languages to low-resource ones. \citeauthor{salza2021effectiveness} \cite{salza2021effectiveness} built a BERT-based model pretrained on multiple languages and transferred it to other languages. CDSC \cite{chai2022cross} adapts a few-shot meta-learning algorithm to transfer code search models trained on widely used languages such as Java and Python to languages with scarce data such as SQL and Solidity. \citeauthor{chen2022transferability} \cite{chen2022transferability} showed that finetuning multilingual pretrained language models on target languages leads to higher performance than training on a single language. These studies differ fundamentally from our work as they transfer models to new languages. We propose methods to transfer models to new projects with the same language. While models cannot perform in new languages, they still do tasks with a reduced accuracy on new projects.
More relevant to this area, we recognize two categories of studies. First, ``Cross-Project Domain Shift and Domain Adaptation in Defect Prediction and Automated Program Repair'' methods. Since the related studies in APR are limited, we expand the scope of related work to defect prediction further. Defect prediction has similar concepts to APR, making them considerable in our literature. Second, ``Semi-supervised Automated Program Repair''. Semi-supervised APR mainly refers to creating APR models for projects that lack labeled data. This is relevant to our bug generator and data synthesis method. The following two sections describe them and review the related studies. 

\subsection{Cross-Project Domain shift and Domain Adaptation on Defect Prediction and Automated Program Repair}
\citeauthor{jin2021cross}\cite{jin2021cross} was one of the firsts who showed that defect prediction methods do not perform similarly on new projects. \citeauthor{nam2013transfer} \cite{nam2013transfer} and \citeauthor{ni2017cluster} \cite{ni2017cluster} confirmed the same issue and proposed a transfer learning method to transfer models to target projects. \citeauthor{ma2012transfer} \cite{ma2012transfer} used a clustering method to train a defect predictor on both target data and part of the source data close to the target. \citeauthor{limsettho2018cross} \cite{limsettho2018cross} used an oversampling method to alleviate class mismatch between different projects. \citeauthor{nguyen2019deep} \cite{nguyen2019deep} proposed a deep domain adaptation method to learn features in target projects for vulnerability detection automatically. \citeauthor{liu2019two} \cite{liu2019two} proposed a source project estimator to select the two closest source projects to the target project and used the method proposed by \citeauthor{nam2013transfer} \cite{nam2013transfer} to create target prediction models. \citeauthor{de2021comparing} \cite{de2021comparing} proposed a transfer learning method to create smelly instances in the target project to address the class imbalance issue. \citeauthor{liu2020cd} \cite{liu2020cd} proposed a cross-domain representation for software programs. First, they used deep learning to create code and domain agnostic representations. Then, they used normal methods to predict vulnerabilities. \citeauthor{jin2021cross} \cite{jin2021cross} proposed a kernel twin support vector machine with DA function to match training data distributions of different projects. 

These studies are all very different from ours in many aspects. First and most important, none of these methods are based on novel transformer-based APR techniques. They are based on traditional methods such as SVM and do not benefit from the natural transferability of transformers. For example, they do not use general pretrained models and instead train new models on target projects from scratch. Therefore, we do not find these studies and their methods helpful for state-of-the-art transformer-based APR methods such as TFix and CodeXGLUE. Second, they are focused on defect prediction task that is easier and different than program repair. In defect prediction, the objective is to predict a binary label showing whether the program is buggy (smelly) or not, while in program repair, the objective is to fix the line. These methods fundamentally differ from the proper DA methods for sequence-to-sequence APR models. For example, the methods that address class imbalance cannot be transferred to our work. More relevant to this work, we found a study that analyzes and addresses domain shifts in program repair. \citeauthor{he2022distribution} \cite{he2022distribution} analyzes the effect of domain shift on two APR methods, including CuBERT \cite{kanade2020learning} and GNN\cite{allamanis2017learning}. They consider models trained on synthetic data, test them on the real data project, and notice the accuracy is dropped. The main difference between our work and theirs is that they do not consider the models trained on realistic datasets such as TFix and CodeBERT. Moreover, they do not propose any domain adaptation method in case a labeled dataset is present. They consider only the case when no labeled data is provided and propose an semi-supervised methods that we will review in the next section.

\subsection{Semi-supervised Automated Program Repair}
As of semi-supervised APR, some studies proposed methods to create synthetic bug datasets \cite{croft2022data}. We highlight that many of them use defect prediction as their downstream task. However, they mainly inject bugs into non-buggy lines. Therefore, the generated bug datasets can also be used for automated program repair. \citeauthor{he2022distribution} \cite{he2022distribution} used methods proposed in \cite{kanade2020learning} to create buggy samples and incorporated them to improve bug detection and repair. The main important difference is that they use rule-based methods for program repair. This makes the bug creation process relatively simple and limited. For example, they can inject only five simple bugs such as variable misuse into codes. Their method does not generalize to other error types. On the other hand, we incorporated NMT-based techniques that produce bugs of 51 error types. DrRepair \cite{yasunaga2020graph} used character-level perturbations to create common compilation errors. Their method cannot create logical errors. \citeauthor{yasunaga2021break} \cite{yasunaga2021break} introduced a dual architecture, including a critic and a breaker for APR. During a repetitive loop, they train a breaker and a critic. The breaker generates bad examples while the critic detects buggy samples. This method is also limited to compilation errors for C and parsing errors for python. It does not aim to generate logical errors inside the code. This method also requires a perfect error detector; otherwise, models can learn to generate incorrect bugs. Our proposed method does not need or rely on bug detectors or compilers. We use CodeXGLUE as one of the APR methods that does not use error detectors. BugLab \cite{allamanis2021self} jointly trains a pair of detector and selector models. Similarly, they cooperate to create hard-to-detect bugs. Their work's limitation is that they focus on four specific bugs, while our method is not limited to specific errors. Besides, they analyze results on bug detection tasks and do not consider APR methods in their literature. They also validate their models only on the manually created dataset. Therefore, It is unclear if their methods create real bugs. SelfAPR \cite{ye2022selfapr} applies perturbations on the AST level and modifies each AST recursively with insert/delete/replace. In this way, they create a buggy dataset and then train repair models on them. Our work is different from SelfAPR in many ways. First, we use machine learning to create buggy codes instead of manually perturbing codes. Manual perturbing is less likely to mimic real word bugs and cover various error types. Second, their work does not benefit from pretrained language models as they solely train models on the target project. We use two-step training to make use of general data as well as target-specific data. Third, their work needs the presence of test cases to select correct buggy samples. On the other hand, no testing system is required in our work. We showed that usage of all generated bugs leads to improvement in accuracy. Fourth, they use a basic transformer to train their APR method and do not compare it with state-of-the-art APR models such as TFix or CodeXGLUE. However, we combine the idea of domain adaptation and bug generation with the novel APR methods. Finally, \citeauthor{patra2021semantic} \cite{patra2021semantic} introduced that extracts bug seeding patterns from bug histories. It finds candidate locations in target codes and applies patterns to create new bugs. Our method directly learns from bug fixes without manual processing, following the paradigm of TFix. However, A comparison between our proposed method and the one of SamSeed can be a future study.

\section{Future Work}
\label{sc:future}
Even though this study focuses on Automated Program Repair, we note that most motivations, DA frameworks, and methods are also applicable to other software engineering tasks. For example, we recognized that two tools of APR (CodeXGLUE and TFix) are sensitive to domain adaptation. However, tools in other software tasks such as fault detection, test generation, and program recommendation can also be sensitive to DA. One interesting future direction is to expand domain adaptation to different areas of software engineering.

In our experiments, we mainly used projects as distinct domains. However, we noted that domains could also be developers. We also showed the feasibility of using lightweight models that can be prepared quickly, loaded into personal systems, and give real-time recommendations. This paves the path to providing developers with effective personalized APR models. Therefore, domain adaptation on the user (developer) level is another interesting future work.

The idea of a unique model that creates synthetic bugs for all types of datasets and models can be a separate interesting future direction. Our method incorporates machine learning and neural machine translation to create generic bugs. Unlike traditional approaches, our method is not limited to specific bug types. However, this idea is not cultivated thoroughly in this study. One can research different models and methods and incorporate more software engineering techniques such as pre-processing, potentially leading to more effective bug synthesis tools. For example, one might design a process to filter out incorrect buggy samples to increase effectiveness. Furthermore, the use-case of synthetic bugs is not limited to APR. Future studies can use them for various software engineering tasks such as fault detection or test generation. They can compare NMT-based bug generation with traditional and static ones (e.g., mutation testing). Accordingly, we recognize it as a separate area of research to be explored in the future.

Finally, another future direction is zero-shot domain adaptation \cite{peng2018zero, kodirov2015unsupervised}. Zero-shot DA refers to the techniques that make the models robust and resistant to domain shifts. For example, they change the model training in a way it only uses \textbf{domain invariant} features. They usually require no extra step in the target domain. This is a massive benefit in terms of efficiency. Consequently, they need no labeled data in the target domain, which is a significant advantage. Adversarial-based DA methods are examples that use GANs \cite{gallego2020incremental, tzeng2017adversarial, wang2020adversarial}. Still, the final comparison of such techniques with our framework is required. Our framework fine-tunes APR models for specific projects, which might be more effective than general DA.

\section{Conclusion}
\label{sc:conclusion}
In this study, we showed that TFix and CodeXGLUE, two pioneer APR techniques, are sensitive to the potential domain shift. We proposed a domain adaptation framework that increases their performance on new projects. We showed the framework is effective, efficient, and reliable. We discussed the differences and advantages of three DA methods,  \full{}, \adapter{} and \curr{}, on TFix and concluded that the final choice depends on the exact conditions. Moreover, we proposed a transformer-based bug generator method that creates synthetic APR datasets. We showed how effectively the synthetic data can replace a missing APR dataset in the DA framework. Finally, we expressed the required modifications to apply our framework to other APR techniques. Finally, We stated exciting points for future work; bringing domain adaptation to other software engineering tasks,  personalizing APR models, or leveraging the bug generator method for related downstream.

-----------
\bibliographystyle{ACM-Reference-Format}
\bibliography{sample-base}

\end{document}